\documentclass[12pt]{iopart}
\usepackage[dvips]{graphicx}
\usepackage{cite}
\begin{document}
\title[Power requirements for sawtooth control in ITER]{Power requirements for electron cyclotron current drive and ion cyclotron resonance heating for sawtooth control in ITER}

\author{IT Chapman$^{1}$, JP Graves$^{2}$, O Sauter$^{2}$, C Zucca$^{2}$, O Asunta$^{3}$, RJ Buttery$^{4}$, S Coda$^{2}$, T Goodman$^{2}$, V Igochine$^{5}$, T Johnson$^{6}$, M Jucker$^{2}$ RJ La Haye$^{4}$, M Lennholm$^{7}$ and JET-EFDA Contributors$^*$}

\address{\small $^{1}$EURATOM/CCFE Fusion Association, Culham Science Centre, Abingdon, Oxon, OX14 3DB, UK \\
\small $^{2}$CRPP, Association EURATOM/Conf\'{e}d\'{e}ration Suisse, EPFL, 1015 Lausanne, Switzerland \\
\small $^{3}$Association EURATOM-Tekes, Aalto University, Department of Applied Physics, P.O.Box 14100 FI-00076 AALTO, Finland \\
\small $^{4}$General Atomics, PO Box 85608, San Diego, CA 92186, USA \\
\small $^{5}$MPI fur Plasmaphysik, EURATOM-Ass D-85748 Garching, Germany \\
\small $^{6}$EURATOM-VR Association, EES, KTH, Stockholm, Sweden \\
\small $^{7}$JET-EFDA, Culham Science Centre, Abingdon, UK OX14 3DB \\
\small $^{*}$See the Appendix of F. Romanelli \emph{et al.}, Fusion Energy 2012 (Proc. 24th Int. Conf. San Diego, 2012) IAEA, (2012)}

\ead{ian.chapman@ccfe.ac.uk}

\begin{abstract}

13MW of electron cyclotron current drive (ECCD) power deposited inside the $q=1$ surface is likely to reduce the sawtooth period in ITER baseline scenario below the level empirically predicted to trigger neo-classical tearing modes (NTMs).
However, since the ECCD control scheme is solely predicated upon changing the local magnetic shear, it is prudent to plan to use a complementary scheme which directly decreases the potential energy of the kink mode in order to reduce the sawtooth period.
In the event that the natural sawtooth period is longer than expected, due to enhanced $\alpha$ particle stabilisation for instance, this ancillary sawtooth control can be provided from $>10$MW of ion cyclotron resonance heating (ICRH) power with a resonance just inside the $q=1$ surface.
Both ECCD and ICRH control schemes would benefit greatly from active feedback of the deposition with respect to the rational surface.
If the $q=1$ surface can be maintained closer to the magnetic axis, the efficacy of ECCD and ICRH schemes significantly increases, the negative effect on the fusion gain is reduced, and off-axis negative-ion neutral beam injection (NNBI) can also be considered for sawtooth control.
Consequently, schemes to reduce the $q=1$ radius are highly desirable, such as early heating to delay the current penetration and, of course, active sawtooth destabilisation to mediate small frequent sawteeth and retain a small $q=1$ radius.
Finally, there remains a residual risk that the ECCD+ICRH control actuators cannot keep the sawtooth period below the threshold for triggering NTMs (since this is derived only from empirical scaling and the control modelling has numerous caveats).
If this is the case, a secondary control scheme of sawtooth stabilisation via ECCD+ICRH+NNBI, interspersed with deliberate triggering of a crash through auxilliary power reduction and simultaneous pre-emptive NTM control by off-axis ECCD has been considered, permitting long transient periods with high fusion gain.
The power requirements for the necessary degree of sawtooth control using either destabilisation or stabilisation schemes are expected to be within the specification of anticipated ICRH and ECRH heating in ITER, provided the requisite power can be dedicated to sawtooth control.
\end{abstract}

\section{Introduction and Background}

Sawtooth control remains an important unresolved issue for baseline scenario \cite{Casper} operation of ITER.
Since the monotonic $q$-profile of such baseline ELMy H-mode plasmas have a large $q=1$ radius, $r_{1}$, with low magnetic shear at the $q=1$ surface, $s_{1}=r_{1}\textrm{d}q/\textrm{d}r$, these plasmas are expected to be unstable to the internal kink mode.
Furthermore, the energetic trapped fusion-born $\alpha$-particles are predicted to lead to significant stabilisation of the internal kink mode \cite{Porcelli1996,ChapmanEPS}, resulting in very long sawtooth periods.
However, such long sawtooth periods have been observed to result in triggering NTMs at lower plasma $\beta$ \cite{Sauter,Buttery,Chapman2010} (where $\beta$ is the pressure normalised to the magnetic pressure) which in turn can significantly degrade plasma confinement.
Consequently, there is an urgent need to assess whether sawtooth control will be achievable in ITER and how much power is required from the actuators at our disposal to attain an acceptable sawtooth period.
Our understanding of internal kink mode stability that underlies sawteeth has improved significantly recently through a combination of analytic understanding, experimental verification and detailed modelling, as reviewed in \cite{ChapmanRev}.
This enhanced understanding now provides a platform from which to make an improved assessment of sawtooth control requirements in ITER.

The two approaches to sawtooth control are to (i) either eliminate or delay the sawtooth crash for as long as possible (stabilisation) or (ii) decrease the sawtooth period to reduce the likelihood of triggering other MHD instabilities (destabilisation).
In ITER, it is foreseen that destabilisation will be employed to keep the sawteeth small and frequent to help flush He ash from the plasma core and to avoid triggering NTMs.
Sawtooth control can be achieved by tailoring the distribution of energetic ions; by changing the radial profiles of the plasma current density and pressure, notably their local gradients near the $q=1$ surface; by rotating the plasma, or changing the rotation shear local to the $q=1$ surface; by shaping the plasma; or by heating the electrons inside the $q=1$ surface.
The primary actuators to achieve these perturbations are electron cyclotron current drive (ECCD), ion cyclotron resonance heating (ICRH) and neutral beam injection (NBI).
The highly localised perturbations to the current density profile achievable with ECCD have been employed to significantly alter sawtooth behaviour on a number of devices. 
By driving current just inside the $q=1$ surface, the magnetic shear at $q=1$ can be increased, and thus result in more frequent sawtooth crashes.
ECCD is foreseen as the primary sawtooth control actuator in the ITER design \cite{IPB2} due to both the highly localised current density that can be achieved when compared to ion cyclotron current drive for instance, and because of the ability to provide real time control of the current drive location by changing the launcher angle of the injected EC beam by using steerable mirrors.
However, complementary control schemes which work via kinetic effects, such as ICRH or NBI, are also useful for sawtooth control in the presence of a population of core energetic particles.

An open question that predicates the assessment of required actuator power level is what an acceptable sawtooth period will be in ITER.
In order to provide some empirical basis for an acceptable sawtooth period in ITER, a multi-machine database has been established and an empirical scaling law derived \cite{Chapman2010}, as described in section \ref{sec:empirical}.

For many years it has been known that trapped energetic particles result in strong stabilisation of sawteeth.
However, passing fast ions can also significantly influence sawtooth behaviour.
For highly energetic ions, the radial drift motion becomes comparable to the radial extent of the kink mode.
In this regime, the kinetic contribution to the mode's potential energy (together with a non-convective contribution to the fluid part of $\delta W$) becomes increasingly important.
When the passing fast ion population is asymmetric in velocity space, there is an important finite orbit contribution to the mode stability. 
The effect of passing ions is enhanced for large effective orbit widths, which is to say, for highly energetic ions (like ICRH or negative ion NBI (N-NBI) in ITER) or for a population with a large fraction of barely passing ions (like ICRH).
Passing fast ions can destabilise the internal kink mode when they are co-passing and the fast ion distribution has a positive gradient from inside to outside $q=1$, or when they are counter-passing, but the deposition is peaked outside the $q=1$ surface.
This mechanism is described in detail in references \cite{Graves2004} and \cite{Graves2009}.
The effect of passing fast ions has been confirmed in NBI experiments in JET \cite{Chapman2007,Chapman2008} and ASDEX Upgrade \cite{Chapman2009} and using He$^{3}$ minority ICRH in JET \cite{GravesAPS}. 
By employing He$^{3}$ minority heating schemes (which are envisaged for ITER ICRF heating), the resultant current drive is negligible \cite{Graves2009,Laxaback}.
Nonetheless, the ICRH can still strongly influence the sawtooth stability, demonstrating that sawtooth control via ICRH can be achieved via a kinetic destabilisation mechanism rather than through local modification of the magnetic shear at $q=1$ \cite{Graves2012,Graves2010}.
Experimental evidence that both ECCD and ICRH control is effective in plasmas with a significant fraction of core energetic particles is given in section \ref{sec:exp}.

Sawtooth stability is strongly influenced by the energetic particles arising from neutral beam injection, ion cyclotron resonance heating and fusion alpha particles.
Previous assessments \cite{ChapmanEPS,Chapman2008} have shown that the N-NBI ions, like the fusion-born alphas, will be strongly stabilising if the resultant distribution function is peaked inside the $q=1$ surface, as is envisioned for baseline scenario operation.
The tools used to model the fast ion distribution functions and their effect on stability are outlined in section \ref{sec:tools}, and the energetic particle distributions are detailed in section \ref{sec:fastions}.

The effect on the sawtooth period from electron cyclotron current drive (the main actuator planned for ITER) is described in section \ref{sec:ECCD} and the effect of ion cyclotron resonance heating is outlined in section \ref{sec:ICRH}.
Finally, an alternative approach to sawtooth control through deliberate stabilisation is discussed in section \ref{sec:stabilisation}.
The conclusions of the study in terms of required power levels are summarised in section \ref{sec:conclusions}.

\section{An Acceptable Sawtooth Period in ITER} \label{sec:empirical}

The neoclassical tearing mode (NTM) is one of the most critical performance-limiting instabilities for baseline scenarios in ITER.
The NTM is a metastable mode which requires a `seed' perturbation in order to be driven unstable and grow \cite{Carrera}, except at very high plasma pressure \cite{Brennan}. 
Various effects have been proposed to prevent NTM growth for small island widths, namely (i) incomplete pressure flattening which occurs when the connection length is long compared to the island width \cite{Fitzpatrick}, (ii) ion polarisation currents arising due to finite orbit width $E \times B$ drifts occurring for ions and electrons across the island region \cite{Wilson,Smolyakov}, which act to replace the missing bootstrap current, and (iii) curvature effects \cite{Kotschenreuther,Lutjens}.
Consequently, NTM growth is generally prohibited in the absence of a sufficiently large seed island in the plasma \cite{LaHaye2000}.
Whilst this seed may be caused by edge localised modes (ELMs) \cite{Buttery2008,Gerhardt} or fast particle-driven fishbones \cite{Gude}, the trigger of most concern is the sawtooth oscillation which typically triggers the NTMs at lower plasma pressures \cite{Gude}.
Many theories have been proposed to explain how the sawtooth crash triggers the NTM, including magnetic coupling \cite{Hegna}, nonlinear `three-wave' coupling \cite{Nave}, changes in the classical tearing stability due to current redistribution inside $q=1$ \cite{Reimerdes,Maget,Koslowski} or changes in the rotation profile resulting in a reversal of the ion polarisation current \cite{Buttery2003} in the modified Rutherford equation governing NTM stability \cite{Sauter1997}.
These models predict that the salient features of the sawtooth crash that should determine the onset of the NTM are the amplitude of the magnetic perturbation, the coupling to the NTM rational surface and any shielding effects such as rotational screening or diamagnetic effects.
However, empirical observation and neural network analysis have determined that the sawtooth period shows far stronger correlation to the triggering of the NTM than the sawtooth amplitude \cite{Buttery2003,Buttery2004,Sauter,Belo,Coda}.

Although the seeding of the NTM by the sawtooth crash remains poorly understood, the empirical observation that deliberately increasing the sawtooth frequency helps to avoid triggering NTMs is now universally accepted and routinely used as a method for NTM mitigation.
The issue of whether a sawtooth period in ITER in the range of 20-50s, as predicted by transport simulations \cite{Jardin,Bateman1998,Waltz,Onjun,Budny2008,Bateman} will avoid triggering NTMs is currently poorly understood, and so a multi-machine empirical scaling is presented here in order to provide some basis for extrapolation and specification of sawtooth control actuators in ITER.
A database of plasma parameters has been established for discharges which exhibit sawteeth, including both crashes which trigger NTMs and those which do not.
This dataset contains details for over 200 shots from nine tokamaks; namely ASDEX Upgrade, DIII-D, HL-2A, JET, JT-60U, MAST, NSTX, TCV and Tore Supra \cite{Chapman2010}.
Naturally, comparing discharges between a large range of tokamaks means that the database contains a wide range of plasma shapes, $q$-profiles, fast ion pressures and fast ion distribution functions, all of which will influence the sawtooth behaviour. 
Similarly, the different $q$-profiles, and thus different magnetic shear between rational surfaces, as well as the different rotation profiles will undoubtedly influence the coupling between the sawtooth oscillations at $q=1$ and NTMs at higher rational surfaces.
The database also incorporates triggered NTMs at three different rational surfaces, $q=4/3, 3/2, 2/1$.
However, retaining such a wide range of plasma parameters means that a ``safe'' operating space, where sawteeth are less likely to trigger NTMs, can be inferred.

The dynamics which determine when the sawtooth crash will occur (in the absence of any deliberate sawtooth control actuators) are predominantly determined by the evolution of the $q$-profile, particularly of the radial position of the $q=1$ surface and the local magnetic shear at $q=1$ \cite{Porcelli1996}.
Since these quantities evolve on the timescale of the resistive diffusion in the plasma core, the sawtooth period has been normalised accordingly.

\begin{figure}
\begin{center}
\includegraphics[width=0.7\textwidth,viewport = 0 5 630 450,clip]{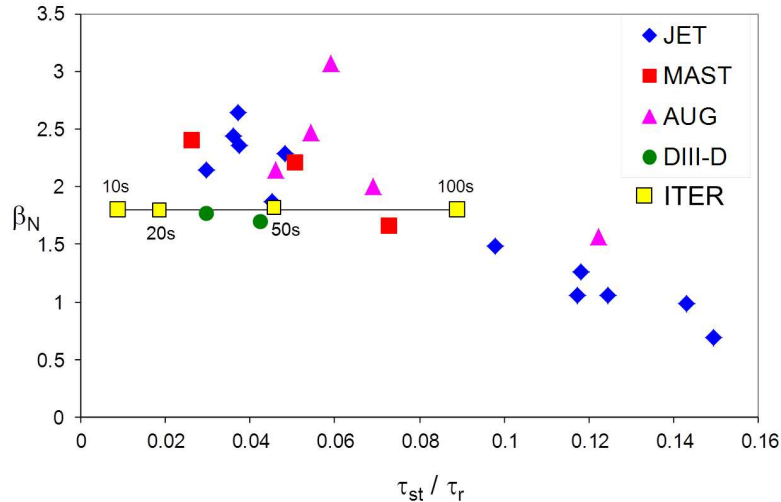}
\end{center}
\caption{$\beta_{N}$ at the NTM onset with respect to the sawtooth period normalised to the resistive diffusion time for ITER-like shape, $q=1$ radius and injected power normalised in a range just above the $P_{LH}$ threshold. For comparison, ITER baseline scenario is indicated with sawtooth period ranging from 10s to 100s. Reproduced with permission from \cite{Chapman2010}.}
\label{fig:trimmed}
\end{figure} 

Notwithstanding the individual constraints of each tokamak, there is a significant scatter in the database meaning that it is difficult to draw any conclusion about the permissible sawtooth period in ITER that avoids triggering NTMs.
Consequently, in order to make a more reliable extrapolation to ITER, a subset of the data has been considered which retains only discharges with ITER like shape ($\delta \in [0.3,0.4]$ and $\kappa \in [1.65,1.85]$), a broad flat $q$-profile with a wide $q=1$ surface ($r_{1}/a \in [0.33,0.45]$) and with auxilliary heating power only slightly above the L-H threshold given by in reference \cite{Martin} ($P_{aux}/P_{LH} \in [1.3,1.7]$) as expected in the ELMy H-mode baseline scenario in ITER \cite{Doyle}.
This reduced database of ``ITER-like'' sawtoothing discharges is illustrated in figure \ref{fig:trimmed}.
It is clear that this subset exhibits a general trend that NTMs are triggered at lower $\beta_{N}$ for longer sawtooth periods with respect to the resistive diffusion time.

Also shown in figure \ref{fig:trimmed} is the range of sawtooth periods that could be expected in ITER. 
A period of 20-50s predicted by transport modelling \cite{Onjun,Budny2008} would lie in the range $\tau_{st}/\tau_{r} \in [0.0178,0.0446]$ which approaches the period at which this empirical extrapolation suggests NTMs would be triggered by the sawtooth crashes at the target plasma pressure of $\beta_{N}=1.8$ in ITER baseline scenario.
However, if the natural sawtooth period is approximately the same as the critical period for triggering NTMs, there is the opportunity to apply control actuators to sufficiently reduce $\tau_{st}$ and avoid NTMs, which would not be the case if the natural period was significantly longer than the critical period.

\begin{figure}
\begin{center}
\includegraphics[width=0.7\textwidth,viewport = 0 5 630 450,clip]{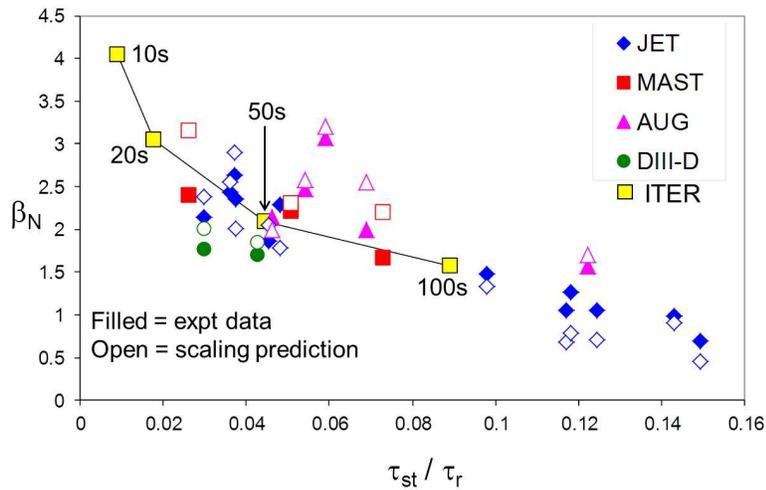}
\end{center}
\caption{The $\beta_{N}$ at which an NTM is triggered with respect to the sawtooth period normalised to the resistive diffusion time (filled symbols) for ITER-like shape, $q=1$ radius and injected power normalised in a range just above the $P_{LH}$ threshold, compared to the critical $\beta_{N}$ predicted by the scaling law (open symbols) from reference \cite{Chapman2010}. For comparison, ITER baseline scenario is indicated with sawtooth period ranging from 10s to 100s. Reproduced with permission from \cite{Chapman2010}.}
\label{fig:trimmed_fit}
\end{figure} 

An empirical scaling law developed from the entire database suggests that the critical $\beta_{N}$ for triggering an NTM by a sawtooth crash in ITER is 2.09 for a sawtooth period of 50s.
Figure \ref{fig:trimmed_fit} shows the critical $\beta_{N}$ for triggering NTMs for the ITER-like subset, compared to the predictions of the derived scaling law, showing good agreement between the two.
Also overlaid is the critical achievable pressure predicted for a range of sawtooth periods in ITER.
At the target operating pressure for ITER ELMy H-mode scenario -- $\beta_{N}=1.8$ -- this scaling law suggests that a sawtooth period of around 70s will be permissible.
It is evident that a sawtooth period in the range of 20-50s predicted by transport simulations is predicted to avoid triggering NTMs at the scenario target operating pressure.
It is also clear that the critical $\beta_{N}$ for NTM onset increases as the sawtooth period is reduced, highlighting the need for provision of sawtooth control actuators.
This scaling law is, of course, only an empirical fitting and not based on any physics model, so its application to future devices should only be for guidance, and certainly not quantitative.
It should be noted that the empirical scaling law derived from the experimental database is primarily for unidirectional NBI-heated plasmas.
Supplementing this database with extra plasmas run at more ITER-relevant low torques would help to clarify whether the rotation plays an important role in mediating the coupling between the sawtooth crash and the NTM onset and is likely to lead to an additional parameter in the scaling law.

\section{Experimental evidence of sawtooth control in the presence of core energetic particles} \label{sec:exp}

When a sawtooth crash occurs in the presence of stabilising fast ions it is often more violent and more likely to trigger NTMs, leading to a degradation in pressure and thus in fusion performance.
Therefore, it is important to demonstrate that both ECCD and ICRH can be used to control sawteeth in the presence of a population of core fast ions.

Whilst ECCD has been shown to control sawteeth effectively for decades, only recently have such demonstrations been replicated in the presence of core energetic particles.
Sawtooth destabilisation of long period sawteeth induced by ICRH generated core fast ions with energies $\geq 0.5$MeV was achieved in Tore Supra, even with modest levels of ECCD power  \cite{Lennholm,Lennholm2009}.
Similarly, ECCD destabilisation has also been achieved in the presence of ICRH accelerated neutral beam injection (NBI) ions in ASDEX Upgrade \cite{Igochine} as well as with normal NBI fast ions in ASDEX Upgrade \cite{Muck} and JT-60U \cite{Isayama}.
More recently sawtooth control using ECCD has even been demonstrated in ITER-like plasmas with a large fast ion fraction, wide $q=1$ radius and long uncontrolled sawtooth periods in DIII-D \cite{ChapmanDIIID}.
As expected from simulation, the sawtooth period is minimised when the ECCD resonance is just inside the $q=1$ surface.
Active sawtooth control using driven current inside $q=1$ allows the avoidance of sawtooth-triggered NTMs, even at much higher pressure than required in the ITER baseline scenario.
Operation at $\beta_{N}=3$ without 3/2 or 2/1 neoclassical tearing modes has been achieved in ITER demonstration plasmas when sawtooth control is applied using only modest ECCD power \cite{ChapmanDIIID}.
Such avoidance of NTMs permitting operation at higher pressure than otherwise achievable by application of core ECCD sawtooth control has also recently been demonstrated in ASDEX Upgrade \cite{ChapmanAUG}.

A major advantage of current drive schemes is that ECCD provides a simple external actuator in a feedback-control loop through the angle of inclination of the launcher mirrors.
Consequently, there has been considerable effort to develop real-time control of the deposition location in order to obtain requested sawtooth periods.
TCV has demonstrated feedback control of the sawtooth period by actuating on the EC launcher injection angle in order to obtain the sawtooth period at a pre-determined value \cite{Paley}.
Recently, fine control over the sawtooth period has been demonstrated on TCV using either `sawtooth pacing' via modulated ECCD with real-time crash detection \cite{Goodman}, or `sawtooth locking', where the sawtooth period is controlled even in the absence of crash detection in a reduced region of duty-cycle versus pulse-period parameter space \cite{Lauret,Witvoet}.
Meanwhile, Tore Supra have implemented a `search and maintain' control algorithm to vary the ECCD absorption location in search of a location at which the sawteeth are minimised; having achieved this, the controller maintains the distance between the ECCD deposition location and the measured inversion radius despite perturbations to the plasma \cite{Lennholm2009}. 

Control of sawteeth by ICRH in the presence of core energetic particles has been widely exploited on JET \cite{Sauter,Eriksson2006,Eriksson2004,Westerhof2002,Mayoral}.
Furthermore, ICRH control has also been demonstrated in plasmas with significant heating power on-axis from neutral beam injection and high $\beta_{p}$, well above the critical threshold for triggering of 3/2 NTMs in the absence of sawtooth control \cite{Coda}.
Subsequently it was noted that the sensitivity of sawtooth destabilisation required accuracy of the resonance position with respect to the $q=1$ surface of less than 0.5\% (ie within 1cm of the $q=1$ surface in JET) \cite{Coda}, far more sensitive than expected from a control mechanism involving a modification of the magnetic shear.
Graves \emph{et al} showed that the sawtooth control mechanism from localised off-axis toroidally propagating waves is due to the radial drift excursion of the energetic ions distributed asymmetrically in the velocity parallel to the magnetic field \cite{GravesAPS}.
This kinetic mechanism results in a deep and narrow minimum in the change of the potential energy when the peak of the passing fast ion distribution is just inside the $q=1$ surface, helping to explain the extreme sensitivity of the sawtooth behaviour to the deposition location of the ICRH waves.
Recent JET experiments using $^{3}$He minority heating (so that the driven current was negligible) on the high-field side just outside the $q=1$ surface lead to a strong destabilisation for counter-propagating waves (-90$^{\circ}$) and a strong stabilisation for co-propagating waves (+90$^{\circ}$) \cite{Graves2010}.
This sawtooth control scheme via affecting the kink mode potential energy has subsequently been demonstrated in H-mode plasmas with significant core heating too \cite{Graves2012}, adding credence to its applicability in ITER.
Finally, real-time control through variation of the ICRH frequency has been attempted on JET \cite{Lennholm2011}, though the frequency variation is much slower than anticipated in ITER \cite{Lamalle}.

\section{Energetic Particle Modelling Tools Used} \label{sec:tools}

\subsection{Modelling the energetic particle distributions}

In order to model the neutral beam fast ion distribution, the \textsc{Transp} \cite{Budny1992} and \textsc{Ascot} \cite{Heikennen,Kurki} codes have been used. \textsc{Ascot} has been used to model the alpha particle population whilst \textsc{Selfo} \cite{Hedin} and \textsc{Scenic} \cite{Jucker,Jucker2} codes have been used to simulate the ICRH distribution. 
Finally, the \textsc{Hagis} drift kinetic code \cite{Pinches} has been employed to study the effect of the various fast ion populations on internal kink stability.
The plasma equilibrium for the ITER baseline scenario is taken from integrated transport modelling using the \textsc{Corsica} code as reported in \cite{Casper}.

The \textsc{Transp}  code was used to simulate the NBI fast ion population since it enables the use of the beam module \textsc{Nubeam} in a convenient, integrated plasma simulation environment.
The \textsc{Nubeam} module is a Monte Carlo package for time dependent modelling
of fast ion species using classical physics.
Multiple fast ion species can be present, due to either beam injection of energetic neutral particles  or as a product of nuclear fusion reactions.
The model self consistently handles guiding center drift orbiting, collisional and atomic physics effects during the slowing down of the fast ions.
In order to reduce the risk of a result dependent entirely upon the prediction of one code, the \textsc{Ascot} code has also been used to simulate the NNBI distribution. 
\textsc{Ascot} \cite{Kurki} is a guiding-centre orbit following Monte Carlo code which integrates the particles' equation of motion in time over a five-dimensional space. Collisions with the background plasma are modelled using Monte Carlo operators allowing an acceleration of collisional time scales and reduced computational time.
The alpha particle markers are initialised by the local $\langle \sigma v \rangle_{DT}$ whereas the beam ions are followed starting from the injector taking into account the beamlet position, direction, beam species, energy, total power, and its bi-Gaussian dispersion.
The ionization cross-section is calculated at each step using the local temperature and density, and analytic fits from \cite{Suzuki}.
In addition to thermal fusion reactions, also fusion reactions between the fast NBI particles and thermal plasma particles are included in the \textsc{Ascot} code. 

The ICRH fast ion populations are simulated using \textsc{Selfo} and \textsc{Scenic}.
The \textsc{Selfo} code \cite{Hedin} determines self-consistently the power absorption and the fast ion acceleration by coupling the global wave solver \textsc{Lion} \cite{Villard} and the Monte-Carlo code \textsc{Fido} \cite{Carlsson}. 
 \textsc{Fido} solves the 3D orbit averaged kinetic equations, including quasilinear ICRF acceleration from the \textsc{Lion} wave field. 
\textsc{Fido} accounts for  guiding centre orbits, including all possible shapes of banana and potato orbits. 
It should be noted that \textsc{Lion} does not include the upshift of the parallel wave number, and therefore \textsc{Selfo} can be used to treat harmonic heating schemes, but not mode conversion. 
A limitation in the \textsc{Fido} code is the assumption of circular flux surfaces. 
To minimise the error caused by this assumption, the ITER equilibrium has been mapped so that the poloidal flux function in the outboard midplane  $\{\psi(R,Z =
Z_{axis}) | R > R_{axis} \}$ is the same in \textsc{Selfo} as in the non-circular ITER equilibrium. 
Furthermore, the ICRH power is normalised so that the power absorbed per resonant ion is the same. 

To reduce the uncertainty in simulating the ICRH distribution with just one code, the \textsc{Scenic} code has also been used.
The \textsc{Scenic} integrated code package \cite{Jucker,Jucker2} takes an equilibrium from \textsc{Animec} \cite{Cooper}, the wave fields and wave numbers from LEMan \cite{Popovich} and iterates with the distribution function evolved by \textsc{Venus} \cite{Fischer,Cooper2007}.
These codes are iterated to form a self-consistent solution which can incorporate anisotropic equilibria in full 3D geometry.
For the equilibrium and wave field computations, a bi-Maxwellian distribution is used for the hot minority, allowing for pressure anisotropy and stronger poloidal dependence of the pressure and dielectric tensor.
Whereas LEMan is limited to leading order FLR effects, and thus to fundamental harmonic without mode conversion, it computes the wave vectors with the help of an iterative scheme, and can therefore treat correctly upshifted wave numbers without the use of a local dispersion relation. 

\begin{figure}
\begin{center}
\includegraphics[width=0.8\textwidth,viewport = 0 0 730 500,clip]{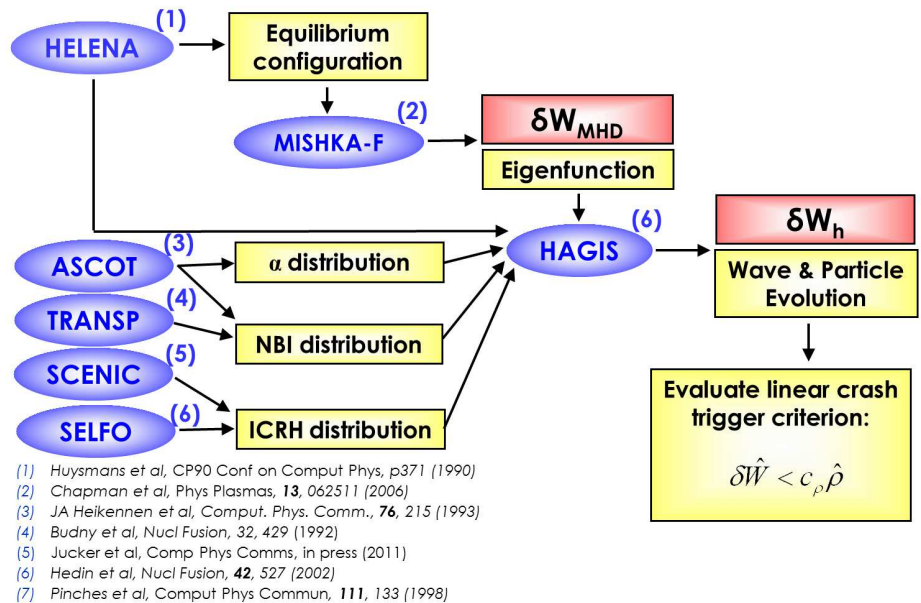}
\end{center}
\caption{A schematic of the suite of codes used to ascertain the stability of the internal kink mode in the presence of energetic particles.}
\label{fig:code_overview}
\end{figure} 

\subsection{Internal kink mode linear stability criteria}

The fundamental trigger of the sawtooth crash is thought to be the onset of an $m=n=1$ internal kink mode.
However, the dynamics of this mode are constrained by many factors, including not only the macroscopic drive from ideal MHD, but collisionless kinetic effects related to high energy particles \cite{Porcelli1991} and thermal particles \cite{Kruskal} as well as non-ideal effects localised in the narrow layer around $q=1$.
A heuristic model predicts that a sawtooth crash will occur when one of three criteria is met \cite{Porcelli1996,SauterVarenna}.
In the presence of fast ions, two conditions are unlikely to be satisfied since the magnetic drift frequency of the hot ions, $\omega_{dh}$, will be large and $\hat{\delta W}$ may have a large positive contribution from $\hat{\delta W}_{h}$.
The change in the kink mode potential energy is defined such that $\delta \hat{W} = \delta \hat{W}_{core} + \delta \hat{W}_{h}$ where $\delta \hat{W}_{core}=\delta \hat{W}_{MHD} + \delta \hat{W}_{KO}$ and $\delta \hat{W}_{KO}$ is the change in the mode energy due to the collisionless thermal ions \cite{Kruskal}, $\delta \hat{W}_{h}$ is the change in energy due to the fast ions and $\delta \hat{W}_{MHD}$ is the ideal fluid mode drive \cite{Bussac}. 
The potential energy is normalised such that $\delta \hat{W} \equiv 4 \delta W/(s_{1} \xi_{0}^{2} \epsilon_{1}^{2} R B^{2})$ and $\xi_{0}$ is the plasma displacement at the axis, $\epsilon_{1}=r_{1}/R$, $R$ is the major radius and $B$ is the magnetic field.
When $\hat{\delta W}$ is large and positive, the mode takes the structure of a tearing mode, which is resistive and can be weakly unstable. 
It is assumed that these drift-tearing modes are stabilised by diamagnetic effects, so do not drive sawtooth crashes.
When the potential energy is sufficient to weakly drive a resistive kink mode, the sawtooth crash is determined by the domain in which the resistive mode can be destabilised, that is to say when
\begin{equation}
-c_{\rho} \hat{\rho} < -\hat{\delta W} < \frac{1}{2} \omega_{*i}\tau_{A}
\label{eq:crashresistive}
\end{equation}
where $\tau_{A}= \sqrt{3}R/v_{A}$ is the Alfv\'{e}n time, $c_{\rho}$ is a normalisation coefficient of the order of unity that determines the threshold at which the mode is considered to result in a sawtooth crash, $\omega_{*i}$ is the ion diamagnetic frequency, $\hat{\rho}=\rho_{\theta i}/r_{1}$ and the poloidal ion Larmor radius is $\rho_{\theta i}=v_{thi}m_{i}/eB_{\theta}$ where $B_{\theta}=\mu_{0}I_{p}/2\pi a$ and $v_{thi}=(kT_{i}/m_{i})^{1/2}$.
In the presence of fast ions, the sawtooth crash is typically triggered by a resistive kink mode when inequality \ref{eq:crashresistive} is satisfied.
However, it should be noted that the crash can still be triggered by an ideal internal kink if the magnetic shear is sufficiently large that the normalisation of $\delta W$ results in a crash.
In ITER, $\rho_{\theta i}$ will be small and $r_{1}$ is expected to be large, meaning that satisfying \ref{eq:crashresistive} by increasing $s_{1}$ alone may not be possible if $\delta W$ is large and positive, so it is prudent to find ways to directly reduce the fast ion stabilisation arising from core energetic particles.

\subsection{Stability modelling} \label{sec:hagis}

The effect of the fast ions on the kink mode stability is tested using the Monte-Carlo guiding centre drift kinetic code \textsc{Hagis} \cite{Pinches}.
The equilibrium is calculated with the static fixed-boundary 2D Grad-Shafranov solver \textsc{Helena} \cite{Huysmans}. 
The stability of this equilibrium is then tested using the linear MHD code \textsc{Mishka} \cite{ChapmanMishka}.
The perturbation and equilibrium are then fed into \textsc{Hagis} together with the distribution functions of fast ions from the modelling described above.
This coupling of numerical codes is illustrated schematically in figure \ref{fig:code_overview}.

\textsc{Hagis} solves the non-linear drift guiding centre equations of motion.
It allows the evolution of a fast ion population to be studied in the presence of electromagnetic perturbations in a toroidal plasma.
The \textsc{Hagis} code has been used extensively for studying the stability of the internal kink mode, successfully replicating experimental signatures of sawtooth behaviour on JET \cite{Chapman2007,ChapmanEPS,Graves2010}, TEXTOR \cite{Chapman2008} and ASDEX Upgrade \cite{Chapman2009,ChapmanPoP}.

\section{Energetic Particle Distributions in ITER} \label{sec:fastions}

\subsection{Core energetic particles}

\begin{figure}
\begin{minipage}{0.4\textwidth} 
\begin{center}
\includegraphics[width=\textwidth,viewport = 0 20 330 400,clip]{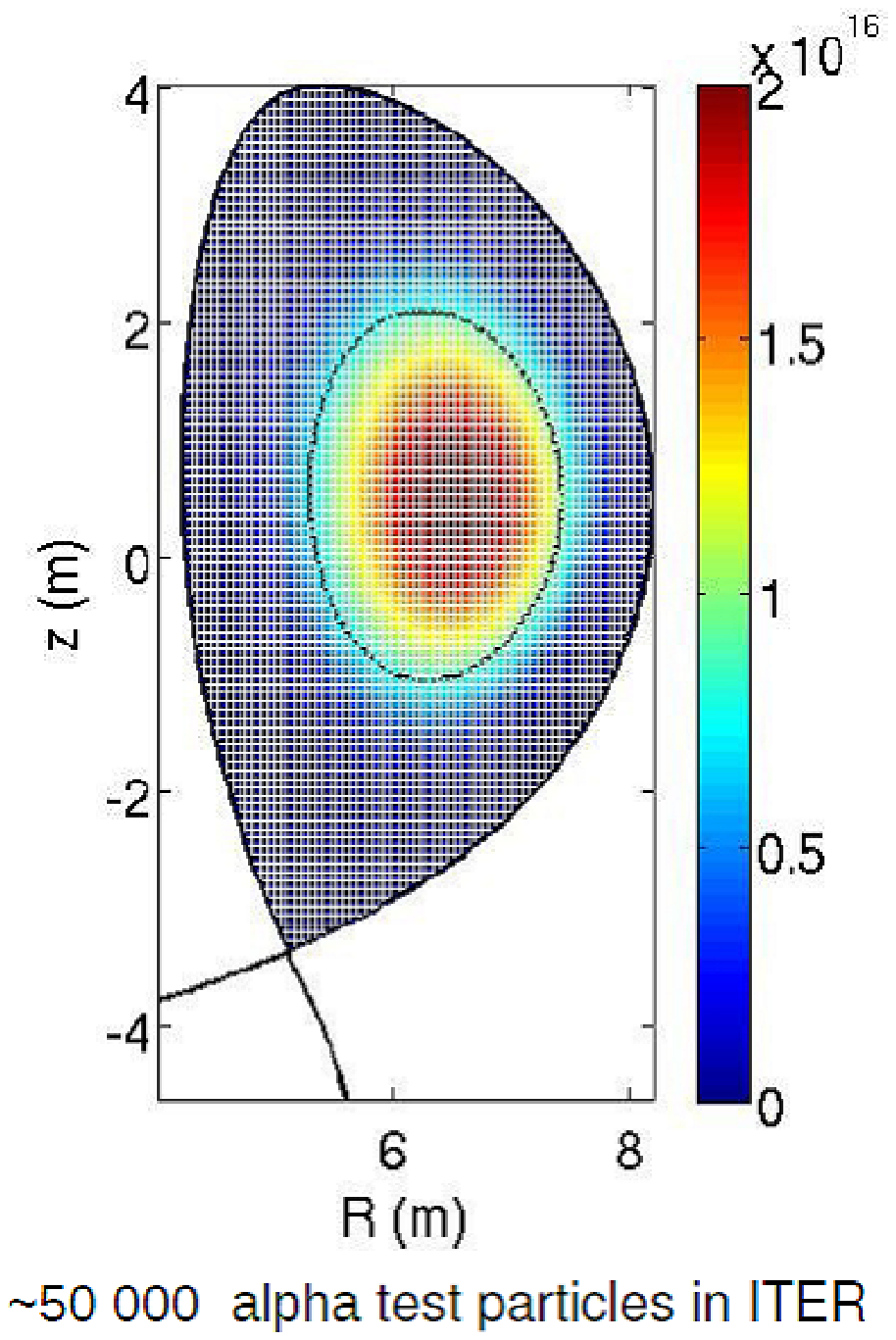}
\end{center}
\end{minipage}
\begin{minipage}{0.6\textwidth}
\begin{center}
\includegraphics[width=\textwidth,viewport = 0 0 380 280,clip]{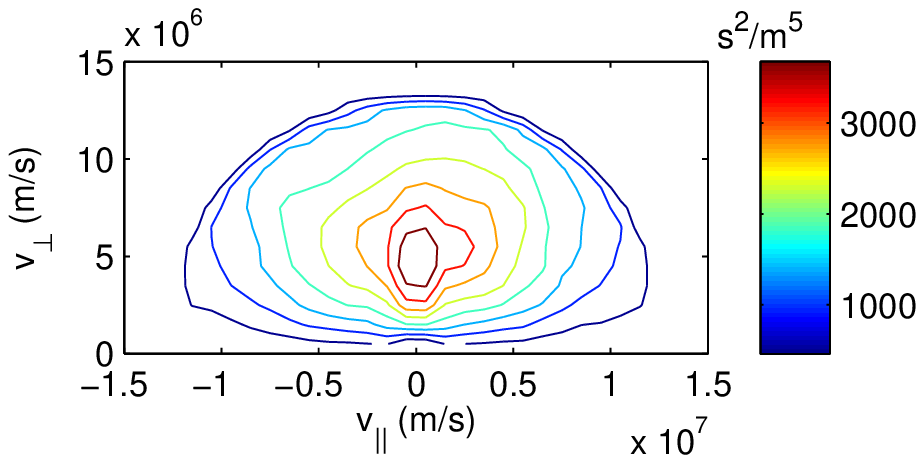}
\end{center}
\end{minipage}
\caption{The birth density, as calculated by \textsc{Ascot}, of the fusion alpha population on (left) an $(R,Z)$ grid  and (right) in perpendicular versus parallel velocity space in ITER, where the velocity distribution is averaged across the minor radius to q=1 in order to represent all fast ions which affect the behaviour of the internal kink mode.}
\label{fig:alpha_ascot}
\end{figure} 

The distribution of alpha particles has been tested with \textsc{Ascot} in the case when there is a 3D equilibrium field due to the presence of ferritic inserts.
The alphas are well confined within $\rho \sim 0.6$ and are approximately isotropic, as seen in figure \ref{fig:alpha_ascot}, where the velocity distribution is averaged across the minor radius to q=1 in order to represent all fast ions which affect the behaviour of the internal kink mode.

In order to penetrate the hot, dense plasmas in ITER, neutral deuterium beam energies of the order of 0.5-1.0MeV are necessary.
In this study, the N-NBI is assumed to consist of 1MeV (D) neutrals from a negative ion-beam system injected in the co-current direction, at a tangency radius of 6m. 
This generates a broad beam-driven current profile with a total driven current of 1.2 MA \cite{Budny2002}.
The beam can be aimed at two extreme (on-axis and off-axis) positions by tilting the beam source around a horizontal axis on its support flange, resulting in N-NBI injection in the range of $Z$ = -0.25 to -0.95 m \cite{ITER}.

\begin{figure}
\begin{center}
\includegraphics[width=0.5\textwidth,viewport = 0 0 530 420,clip]{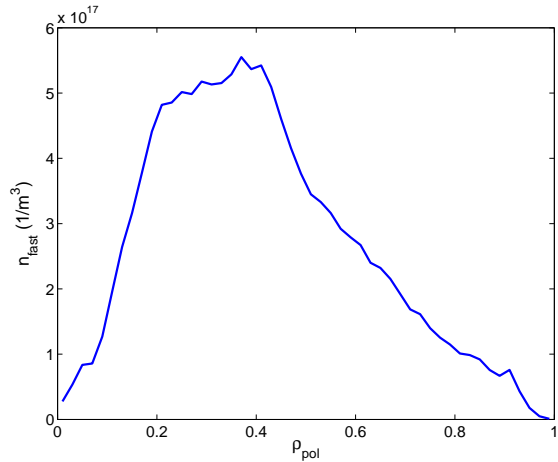}
\end{center}
\caption{The radial distribution of the fast ions resultant from off-axis NNBI injection as predicted by the \textsc{Ascot} code.}
\label{fig:nnbi_radial}
\end{figure}

\textsc{Transp} and \textsc{Ascot} simulations have been carried out to predict the fast ion distribution function due to the N-NBI when it is aimed either on- or off-axis \cite{Budny2002}.
The off-axis fast ion population is peaked at approximately $r/a=0.22$, as seen in figure \ref{fig:nnbi_radial}.
This fast ion population is strongly passing. 
The current driven by the neutral beams results in the $q=1$ surface being slightly closer to the magnetic axis than when on-axis NBI is applied.

\subsection{Ion cyclotron resonance heating}

The application of $^{3}$He minority heating in baseline scenario with the resonance on-axis, slightly off-axis and near mid-radius have been simulated for a range of different minority concentrations.
The phasing of the antenna has also been investigated, and it is found that the inward pinch with +90$^{\circ}$ phasing enhances the on axis fast ion pressure.
For the case with 20MW injected on-axis and minority concentration of $n_{^{3}He}/n=0.01$ simulated with \textsc{Selfo}, around 70\% of the power absorbed goes into heating the $^{3}$He ions (around 7MW).
The off-axis resonance has also been simulated in order to generate a strong radial gradient in the asymmetry of the passing fast ion distribution near the $q=1$ surface necessary for sawtooth destabilisation.
Whilst the far-off-axis heating gives rise to a low power per particle and no highly energetic tails in the distribution, it does nonetheless incur fast ion distributions capable of affecting internal kink stability.

The orbit width effects upon which the internal kink destabilisation mechanism are predicated \cite{Graves2009} are much smaller in ITER than in JET.
In ITER, with $^{3}$He minority heating at 52MHz (ie resonance 0.16m from the magnetic axis) and toroidal field $B_{T}=5.3$T, 1MeV ions have an orbit width $\Delta_{r}/a = 0.06$, whereas for comparison, 1MeV ions in JET with $^{3}$He minority at $B_{T}=2.75$T (as per experiments in reference \cite{Graves2010}) have an orbit width of $\Delta_{r}/a = 0.25$.
The fast ion effects are the only way in which the ICRH can contribute to internal kink stability since the strong electron drag means that the change in the magnnetic shear due to ICCD will be negligible \cite{Laxaback}. 

\begin{figure}
\begin{minipage}{0.3\textwidth} 
\begin{center}
\includegraphics[width=\textwidth]{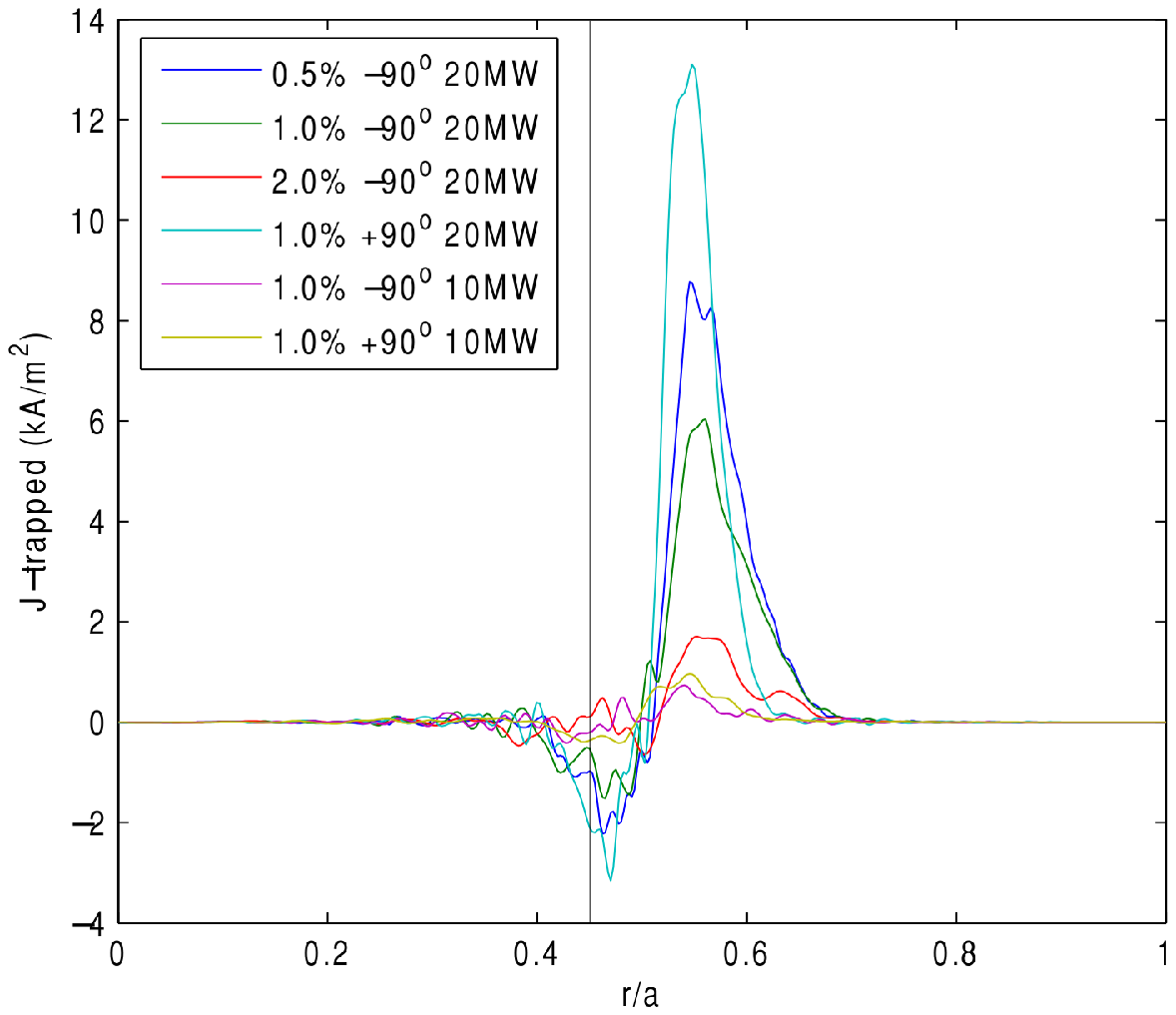}
\end{center}
\end{minipage}
\begin{minipage}{0.3\textwidth}
\begin{center}
\includegraphics[width=\textwidth]{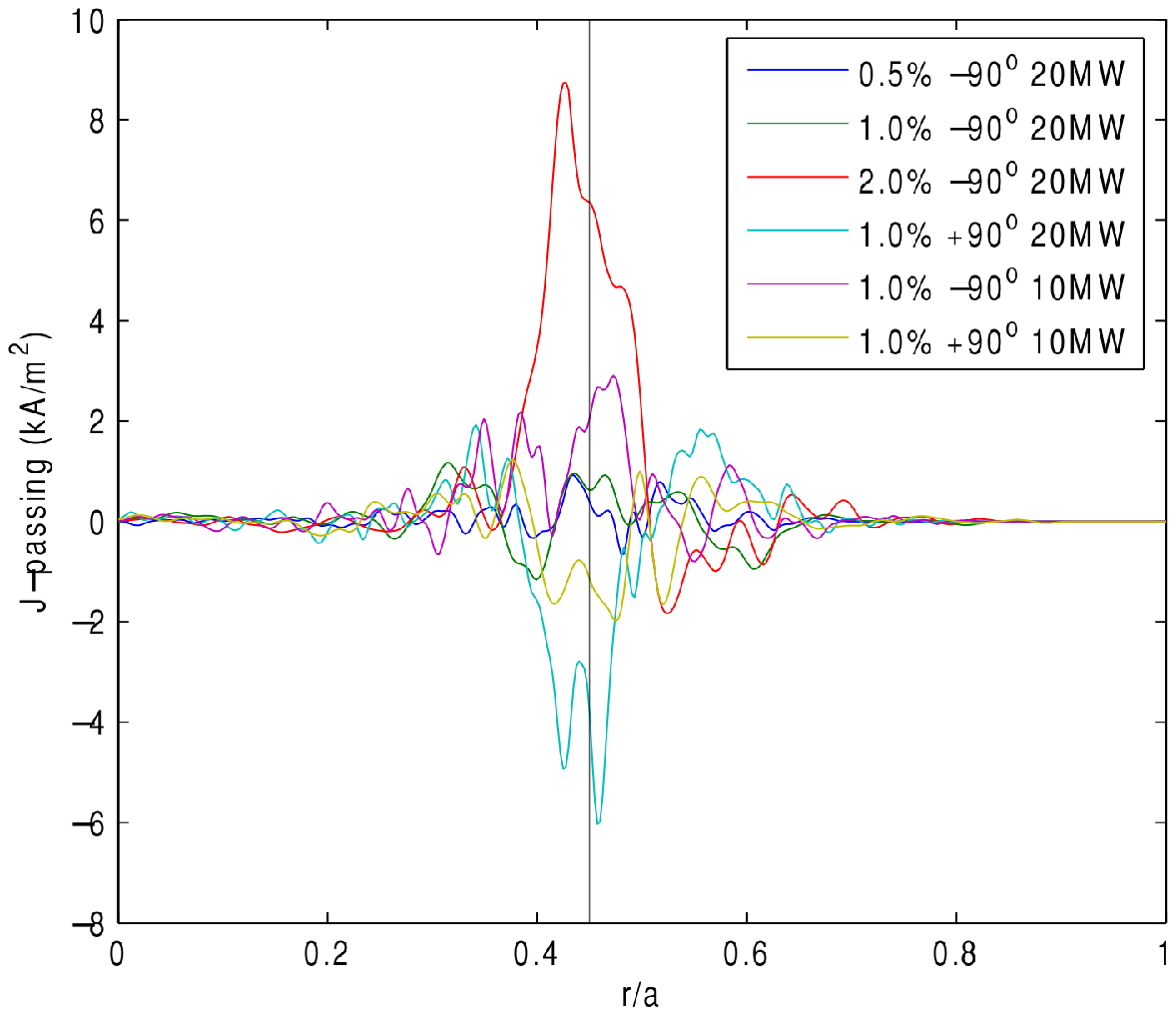}
\end{center}
\end{minipage}
\begin{minipage}{0.3\textwidth} 
\begin{center}
\includegraphics[width=\textwidth]{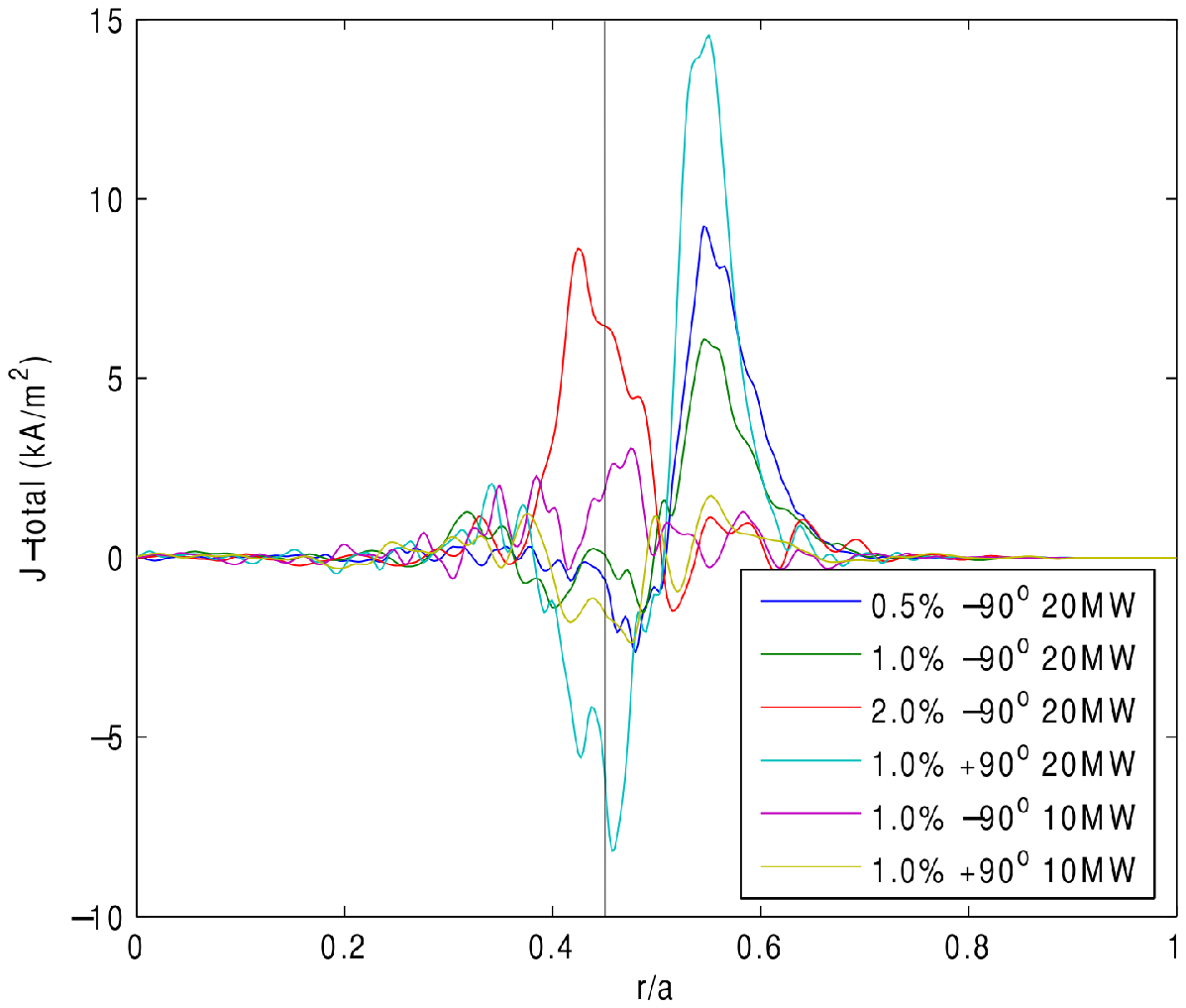}
\end{center}
\end{minipage}
\caption{The fast ion currents arising from ion cyclotron resonance heating of $^{3}$He minority near the $q=1$ surface on the low-field side for different minority concentrations predicted by \textsc{Selfo}. The position of the $q=1$ surface is shown by a vertical line.}
\label{fig:fastioncurrents}
\end{figure}

Figure \ref{fig:fastioncurrents} shows the passing and trapped contributions of the deduced flux-averaged fast ion current density predicted by \textsc{Selfo} as a function of minor radius for ITER at full magnetic field when the ICRH frequency is chosen so that the resonance layer is on the low-field side near the $q=1$ surface.
The current drive predicted by \textsc{Selfo} is smaller than in \textsc{Scenic} and the passing current in \textsc{Selfo} and \textsc{Scenic} is different. 
This may be due to differences in the trapped-passing boundary (the boundary between deeply trapped orbits and passing orbits near low-field side stagnation) in \textsc{Selfo} and \textsc{Scenic}, such that some passing orbits in \textsc{Scenic} are counted as trapped in \textsc{Selfo}. 
Also, the  precession of deeply trapped ions and stagnation passing ions is different. 
A similar benchmark between \textsc{Selfo} and \textsc{scenic} and the iterative coupling between the \textsc{Aorsa} wave field code \cite{Jaeger} and the Fokker-Planck CQL3D code \cite{Harvey} (which neglects orbit widths) was reported in reference \cite{Chapman2011}.


\section{Sawtooth control with ECCD in ITER} \label{sec:ECCD}

The ITER electron cyclotron heating and current drive system consists of up to 26 gyrotrons operating at 170GHz and delivering 1-2 MW each, for a nominal injected power into the plasma of up to 24 MW \cite{Omori}. 
The system has two types of antennas to inject the power into the plasma: the equatorial launcher (EL), which occupies one port in the equatorial plane, and the upper launcher (UL), occupying four ports in the upper plane. 
The EL is designed to access the inner half of the plasma, encompassing all physics applications other than NTM stabilization, including current profile tailoring for steady-state operation, central heating to assist the transition from L- to H-mode and control of the sawtooth instability.
The UL provides a more focussed, peaked driven current density profile, ideal for control of instabilities, including sawteeth after a design modification allowed the UL to access towards the plasma core \cite{Ramponi}. 
In order to access the region from $0.4 < \rho < 0.5$ with the EC power, the access range of the Upper Steering Mirror (USM) and Lower Steering Mirror (LSM) is spread out.
This forms essentially three access zones from the UL: an inner zone accessible with the USM (13MW, 16 beams), an overlap zone accessible with both the USM and the LSM (therefore up to 20MW, 24 beams) and an outer zone accessible with the LSM (13MW, 16 beams). 
Using this method, the overall access region from the UL is increased from about $0.51 < \rho < 0.87$ to about $0.3 < \rho < 0.86$ \cite{Omori,Ramponi}. 

\subsection{Modelling the change in magnetic shear}

The effect of local EC heating on the $q$-profile has been modelled with the \textsc{Astra} transport code \cite{Pereverzev}, which solves a reduced set of 1-D equations for the evolution of the electron and ion temperatures, the helium density and the poloidal magnetic flux. 
The equilibrium is self-consistently calculated with a 2D fixed-boundary solver. 
The electron density is kept fixed and the impurity densities are assumed to be known fractions of it.
The deuterium and tritium densities are determined from quasi-neutrality assuming they are equal, and the effective charge profile is uniform. 
The electron and ion heat diffusion coefficients are normalized to achieve a thermal confinement improvement $H_{(y2,98)} \sim 1$ for the standard ELMy H-mode.
The neoclassical conductivity and the bootstrap coefficients are evaluated by formulas obtained by solving the Fokker-Planck equation with the full collision operator \cite{Sauter1999,Sauter1999b}. 
The Neutral Beam (NB) components are self-consistently evaluated with a Fokker-Planck subroutine which calculates the separate NB contributions to the electrons and ions. 
The EC power density and current driven profiles are evaluated by the beam tracing \textsc{Gray} code \cite{Gray}.

\begin{figure}
\begin{center}
\includegraphics[width=0.5\textwidth,viewport = 0 0 400 500,clip]{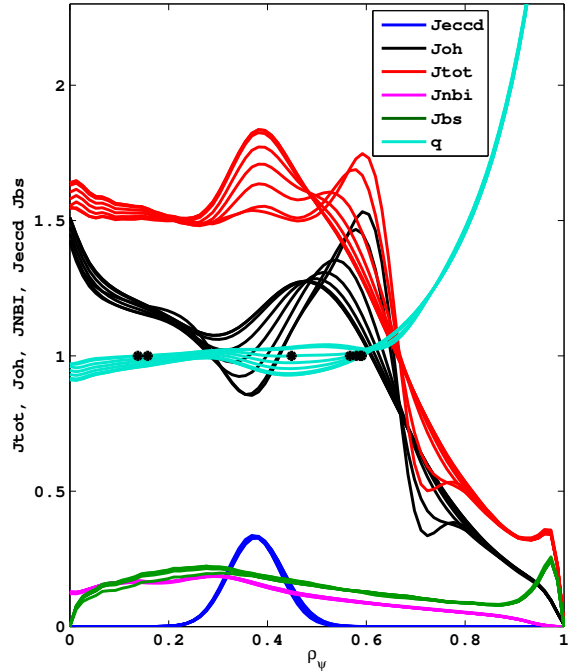}
\end{center}
\caption{The current density and safety factor radial profiles as a function of normalised toroidal flux, $\rho_{\psi}$ evolving in time over one sawtooth period for ITER baseline scenario with 3 equatorial launchers used to provide co-ECCD just inside $q=1$ as modelled by \textsc{Gray}. The profiles are shown through a sawtooth period of 26.6s with the first profile just before the crash and the last profile just after the next sawtooth crash. The black dots represent the position of $q=1$ for each profile.}
\label{fig:currentprofile}
\end{figure}

The ECCD components provided as an input to \textsc{Astra} are Gaussian profiles with amplitude, width and total EC current derived from averaged values output from \textsc{Gray}.
Figure \ref{fig:currentprofile} shows the time evolution of the electron cyclotron driven current during the sawtooth ramp phase together with the Ohmic current, the beam current and the bootstrap current predicted by \textsc{Astra} when the equatorial launchers are used to provide 13.3MW of ECCD just inside the $q=1$ surface.

\begin{figure}
\hspace{-0.5cm}
\begin{minipage}{0.48\textwidth} 
\begin{center}
\includegraphics[width=\textwidth,viewport = 0 5 400 500,clip]{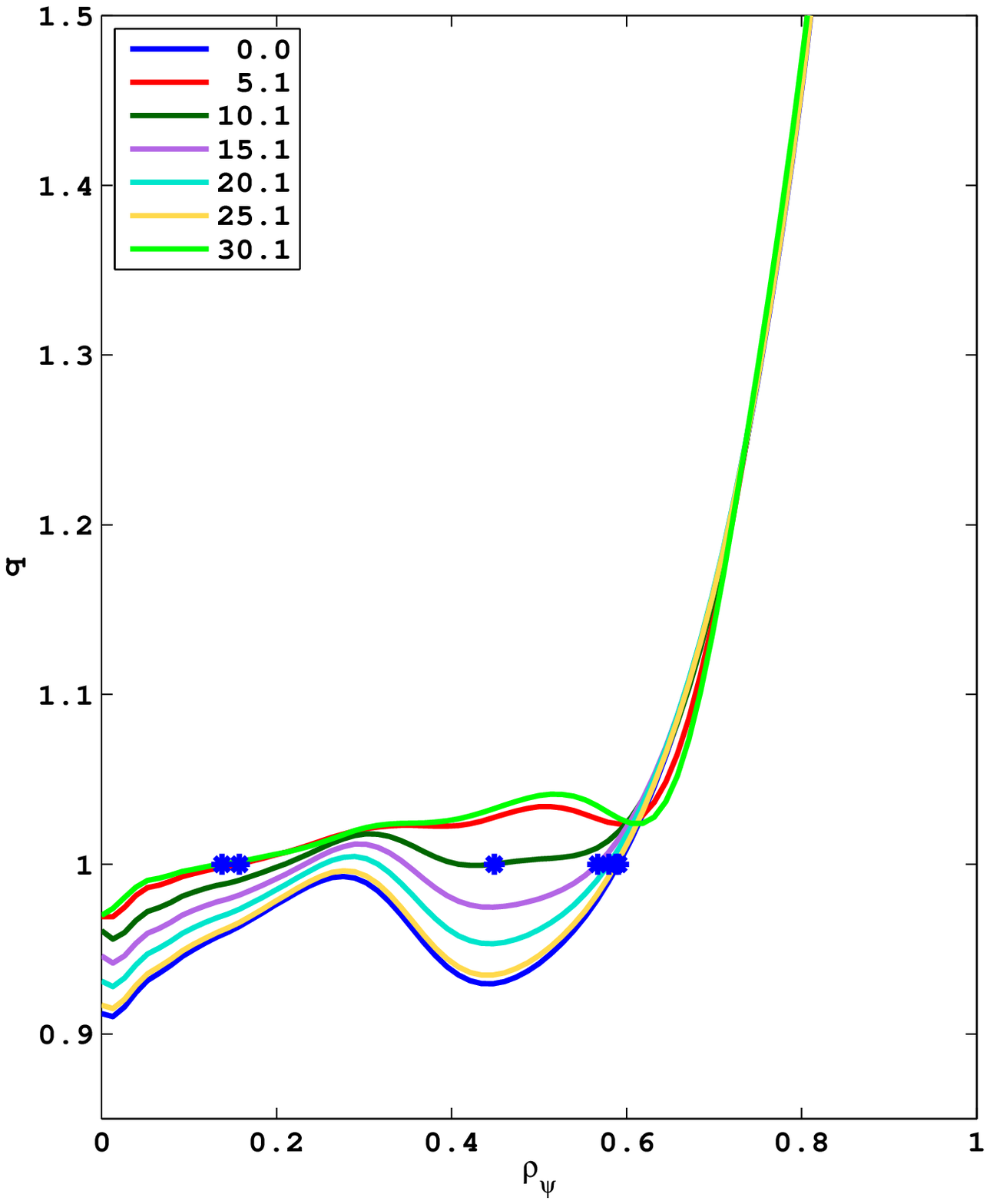}
\end{center}
\end{minipage}
\hspace{0.5cm} 
\begin{minipage}{0.48\textwidth}
\begin{center}
\includegraphics[width=\textwidth,viewport = 0 5 400 500,clip]{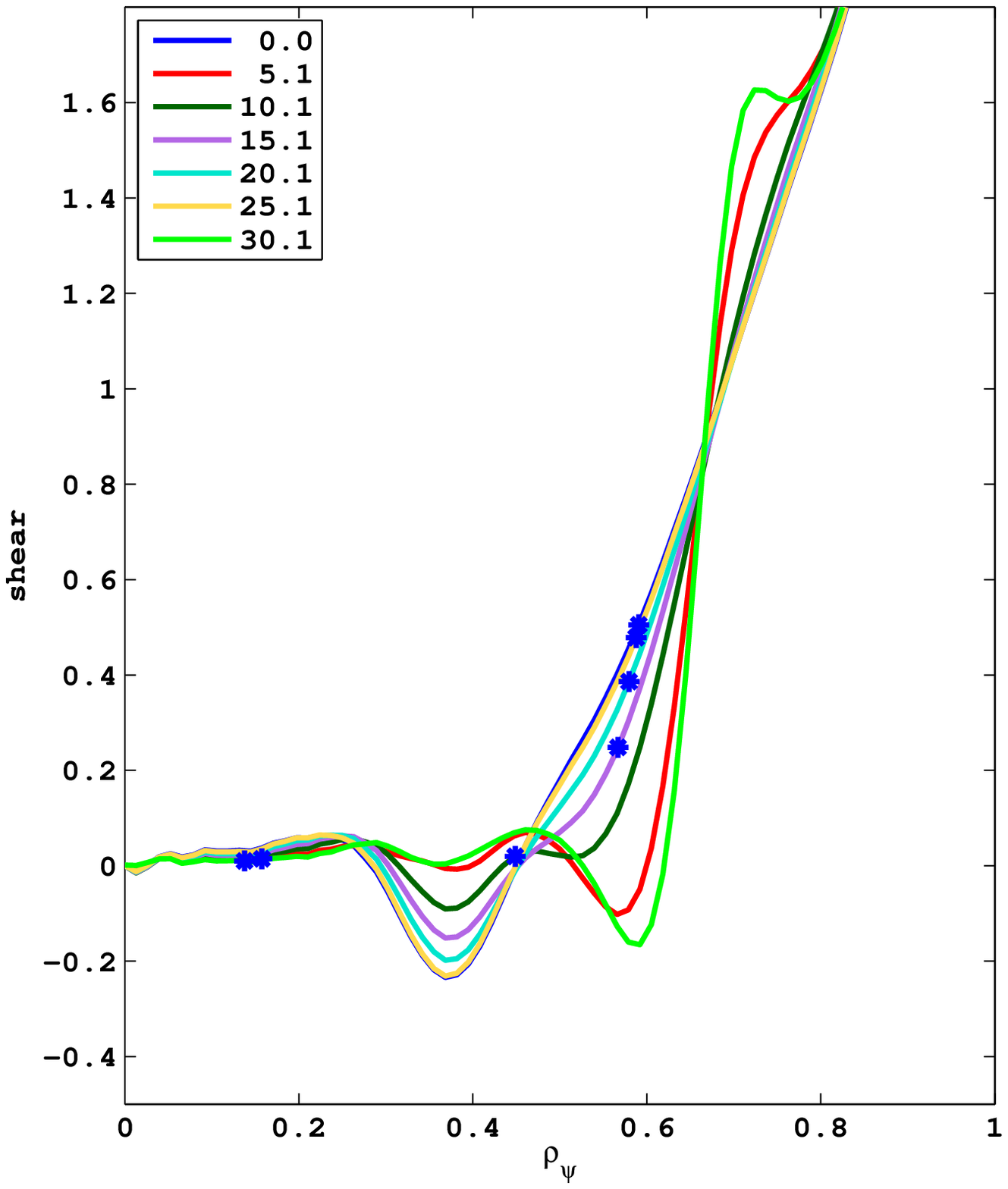}
\end{center}
\end{minipage}
\caption{The same simulation as figure \ref{fig:currentprofile} but showing the evolution of the (left) safety factor profile and (right) magnetic shear profile as a function of normalised toroidal flux, $\rho_{\psi}$ for seven timeslices over 30.1s (compared to $\tau_{ST}=26.6$s) in ITER baseline scenario with 3 equatorial launchers providing 13MW of ECCD near $q=1$ as modelled by \textsc{Gray}. The profiles are shown through a sawtooth period of 26.6s with the first profile just before the crash and the last profile just after the next sawtooth crash. The dots represent the position of the $q=1$ surface. The magnetic shear builds up on a timescale of the order of 15s.}
\label{fig:change_in_q}
\end{figure}

Figure \ref{fig:change_in_q} shows the $q$-profile and the magnetic shear profiles as they evolve in time when the ECCD is applied inside $q=1$.
It is evident that the shear at $q=1$ (here assumed at a radius $\rho_{1} = 0.48$) increases in time after the sawtooth crash (which happens every 26.6s in this simulation) due to the electron cyclotron driven current.
Without ECCD, the value of the magnetic shear at $q = 1$ is 0.15, which is a typical value expected at sawtooth crashes in present experiments in the absence of fast particle stabilization. 
By depositing co-ECCD inside or outside the $q = 1$ radius, the shear at $q = 1$ spans a rather large range, as it changes from just above 0 to 0.4. 

The change in the magnetic shear generated by the application of ECRH has been tested for a range of different launcher configurations \cite{GravesFST}.
When the deposition location is far outside the $q = 1$ surface, there is no significant effect on the shear at the $q = 1$ location: the $s_1$ value stays approximately constant around 0.15.
With the deposition just outside $q=1$, the $s_{1}$ value drops close to zero, and then as the resonance moves inward, the shear rapidly increases and stays constant at approximately $s_{1}=0.4$, even for very on-axis heating.
It should also be noted that the $q = 1$ radius changes rapidly as the ECCD deposition moves across its initial position, meaning that the deposition needs to be adjusted in real time in order to follow the $q = 1$ radius and allow optimum sawtooth destabilization.
That said, the significant increase in $s_{1}$ achievable with core ECCD means that good shear control can be achieved irrespective of the exact resonance provided the EC deposition is inside $q=1$, relaxing requirements on the real-time control system.

\begin{figure}
\begin{center}
\includegraphics[width=0.7\textwidth,viewport = 0 0 400 300,clip]{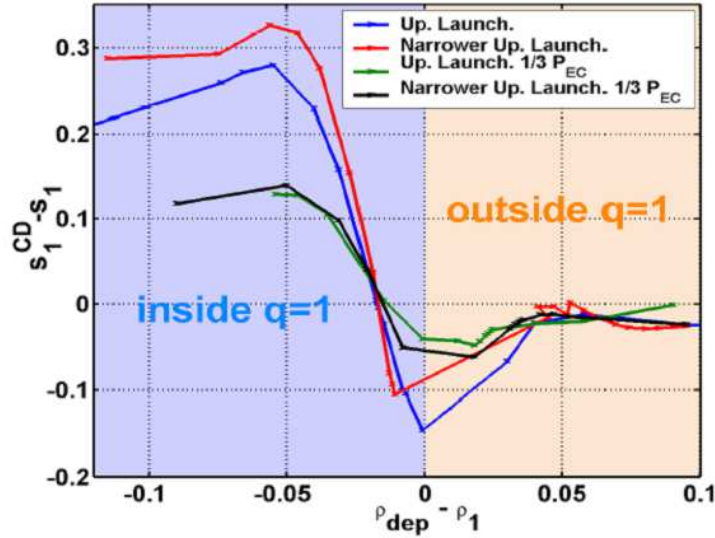}
\end{center}
\caption{The change in the magnetic shear at $q=1$ with respect to the value without ECCD as a function of the deposition position with respect to the radial position of the $q=1$ surface for a range of EC launcher configurations.}
\label{fig:shear_comp}
\end{figure}

Figure \ref{fig:shear_comp} shows the difference of $s_1$ from the original value without ECCD as a function of $\rho_{dep}$ minus the $q = 1$ radius of the case without ECCD; the light blue shaded region corresponds to deposition inside $q = 1$ and the yellow region to deposition outside $q = 1$. 
With deposition inside $q = 1$, $s_1^{CD} - s_1^0$ increases, indicating that a sawtooth crash is likely to be triggered more rapidly and $\tau_{ST}$ decreases (and vice-versa with deposition outside $q = 1$).
In fact, the shear increases when $\rho_{dep} - \rho_{1} \leq -0.02$, rather than strictly anywhere inside $q=1$, though this does not affect the real-time control scheme.
The equilibrium modelling shows that the sawtooth destabilization should be somewhat easier to obtain than stabilization, because the radial extent inside $q = 1$ at which one can deposit co-ECCD and still obtain a significant shear increase is large, whereas one has to be very well localized around a specific region outside $q = 1$ to obtain a significant decrease in $s_1$ and thus have a chance to stabilize sawteeth. 

\subsection{Modelling the effect on the sawtooth period}

In order to model the nonlinear sawtooth period, the \textsc{Astra} transport code includes a heuristic model for when a sawtooth crash will occur, as described in reference \cite{SauterVarenna}. 
The sawtooth period in ITER is predicted to be considerable, due to the influence of $\alpha$-particle stabilization \cite{Chapman2011}.
Since the sawtooth period is related to the free parameter of the model, $c_{\rho}$, this has been chosen to provide $\tau_{ST} = 40s$ for the reference case without any additional EC power, which is a lower bound of the sawtooth period for triggering NTMs predicted from empirical evidence in section \ref{sec:empirical}.
The corresponding value for $c_{\rho}$ is 4.3. 
If the free parameter in the Porcelli model is taken to be $c_{\rho}=1$ (as originally in the model \cite{Porcelli1996}), the sawtooth period approaches 200s and the safety factor on axis drops very low, making these predictions unreliable.

\begin{figure}
\begin{center}
\includegraphics[width=0.7\textwidth,viewport = 0 0 300 210,clip]{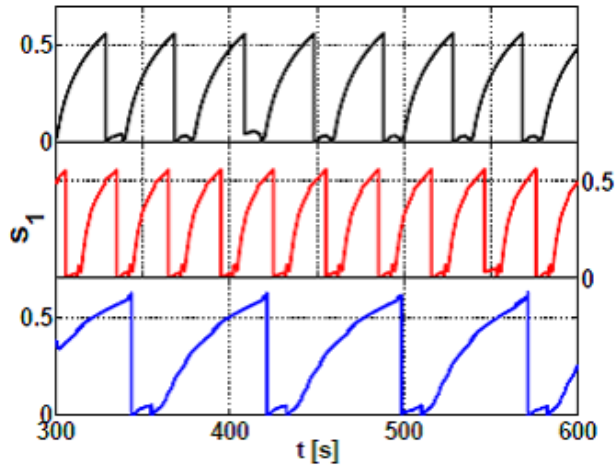}
\end{center}
\caption{The time evolution of the magnetic shear at $q=1$ for the reference case without ECCD (top panel), with ECCD using real-time feedback inside the $q=1$ surface (middle panel) and outside $q=1$ (bottom panel) with off-axis co-ECCD driven by the upper launcher.}
\label{fig:shear_time}
\end{figure} 

The results obtained with 13.3MW of co-ECCD driven from the upper launcher are shown in Figure \ref{fig:shear_time} \cite{Zucca,Zucca2}. 
The plot shows the time evolution of the $s_1$ value in the case with no EC injection (top plot), co-ECCD deposited just inside $q = 1$ (middle plot) and just outside $q = 1$ (bottom plot). 
The sawtooth period can be easily estimated from this plot as $\tau_{ST} = 40$s(top plot), $\tau_{ST} = 30$s (middle plot) and $\tau_{ST} = 70$s (bottom plot).

\begin{figure}
\begin{center}
\includegraphics[width=0.7\textwidth,viewport = 0 5 400 300,clip]{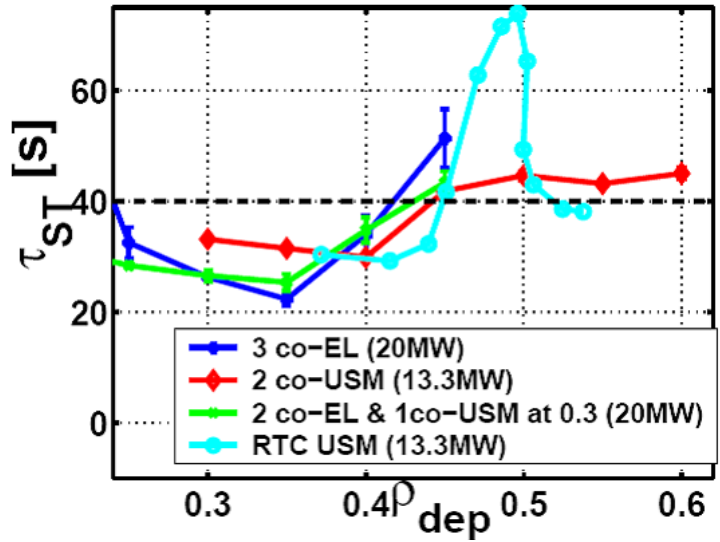}
\end{center}
\caption{The predicted sawtooth period using different combinations of ECCD launchers and powers when the deposition is at different radii as predicted by \textsc{Astra} compared to the simulated period with ECCD (dashed line).}
\label{fig:tau_ECCD}
\end{figure} 

Different launcher variations have been tested for their efficiency in destabilising the sawteeth.
Figure \ref{fig:tau_ECCD} shows the sawtooth period as a function of the radial deposition location, $\rho_{dep}$, of the injected co-ECCD for different mixtures of EC power from the equatorial launcher or the upper launcher. 
The most efficient design uses 20MW of EC power from the equatorial launcher.
In this case, the sawtooth period can be reduced from the reference case of 40s to 23-24s with $\rho_{dep}=0.35$ meaning that the ECCD also leads to efficient heating of the core and minimal impact on the fusion gain, $Q=P_{fusion}/P_{aux}$ \cite{Kirneva}.
A combination of co-ECCD driven by 2 rows of the equatorial launcher (13.3MW) and the remaining power driven by the upper launcher, at a fixed location, can also decrease the sawtooth period down to less than 30s.

The degraded control in both destabilizing and stabilizing the sawteeth by using only the upper launcher can be ameliorated with a real-time control (RTC) algorithm, through which the deposition location is recalculated every time step by the simple formula: $\rho_{dep} = \rho_{1} + \eta w_{CD}$, where $\eta$ is a real-time control parameter that was scanned between -2 and +2, and $w_{CD}$ is the width of the Gaussian ECCD profile.
With real-time feedback controlling the $\rho_{dep}$, the sawtooth period can be increased up to 70s, i.e. more than 50\% increase on the fixed deposition case. 
Note that the time evolution of the value of the magnetic shear at the $q = 1$ surface, shown in Figure \ref{fig:shear_time}(b) and (c), result from the RTC simulations and correspond to a RTC parameter $\eta = 0.75 (\tau_{ST} = 30s)$ and $\eta = 0.25 (\tau_{ST} = 75s)$ respectively.

The \textsc{Astra} modelling suggests that the natural sawtooth period can be reduced by around 30\% by using 13.3MW from the equatorial launchers. 
Real-time control further enhances the ability of the ECCD to destabilise the kink mode and increase the sawtooth frequency.
13.3MW was assumed for control in order to leave more than 5MW available for NTM control if required (references \cite{Sauter2010,LaHaye2009} suggests that relatively low ECCD power is likely to be sufficient for NTM island suppression in ITER).
This assumption also allows for some margin if the ECCD effect is not as efficient as simulated.
Assuming that the natural sawtooth period is approximately 50s as predicted by various transport simulations \cite{Jardin,Bateman,Waltz,Onjun,Bateman1998}, then a reduction of $\sim 30\%$ to 35s is likely to avoid triggering NTMs according to the empirical scaling presented in section \ref{sec:empirical}.
However, the largest uncertainty is what the natural sawtooth period will be.
The reference of 40s can be justified by scaling by resistive diffusion time from the 1s monster sawtooth crashes in JET, and further lengthening to account for the stabilising effect of the alphas by scaling the period in proportion to $\delta W_{\alpha}$ in ITER with respect to that in JET.
Furthermore, the value of $c_{\rho}$ results in a crash when the magnetic shear at $q=1$ is in the range 0.5-0.6, which is in line with typical empirical evidence on a number of devices when active control is applied in plasmas with a large fast ion fraction of the total pressure.

Whilst the requirements of 13.3MW from the equatorial launcher, preferably in real-time control, may be sufficient, the next two sections detail a further ancillary control scheme using ICRH to aid the ECCD destabilisation (section \ref{sec:ICRH}) or an alternative scheme of sawtooth stabilisation (section \ref{sec:stabilisation}).

\section{Sawtooth control with ICRH} \label{sec:ICRH}

The largest risk to controlling sawteeth only with ECCD is that this control scheme works through modification of the magnetic shear profile alone and does not directly influence the free energy to drive the kink mode.
Consequently, it is recommended that a prudent approach will be to consider a complementary scheme which directly affects $\delta W$ and can ideally compete with the stabilisation afforded by the presence of the alpha population.
In order to assess the effect of the ICRH born energetic particles on the stability of the internal kink, the distribution of fast ions simulated by \textsc{Selfo} and \textsc{Scenic} have been fed as a Monte Carlo set of markers into the \textsc{Hagis} code, as described in section \ref{sec:hagis}, and the $\delta W_{ICRH}$ calculated.
This is then compared to the potential energy contributions from the NNBI, the fusion-born alphas and the thermal ions and fluid drive to assess the linear stability of the kink mode.
Whilst such a linear assessment cannot be used to infer the sawtooth period, it qualitatively provides insight into the applicability of ICRH as a control tool, and the ratio of these contributions can be considered as a guide to its efficacy.
Applying the ICRH off-axis means deposition at lower temperature at mid-radius and therefore shorter slowing down, which makes it more difficult to generate as many fast particles.
This means that there are not very energetic tails to the distribution, and the absence of very fast ($>$10MeV) particles means that the finite orbit width effects are diminished.
That said, however, the ICRH ions do still have a relatively strong impact on the internal kink stability.

\begin{figure}
\begin{center}
\includegraphics[width=0.7\textwidth,viewport = 0 0 600 390,clip]{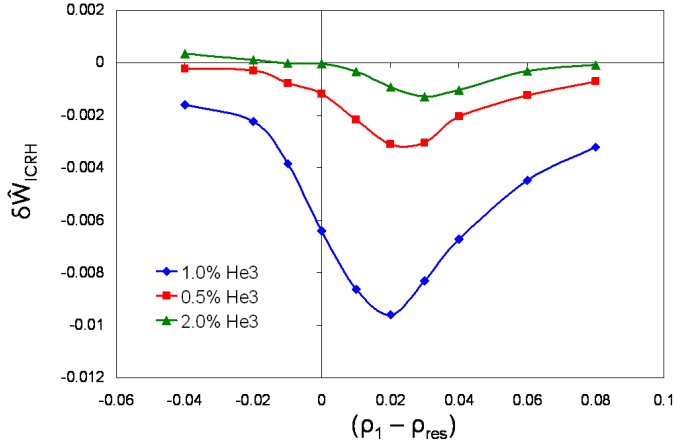}
\end{center}
\caption{$\delta W_{ICRH}$ as the position of the $q=1$ surface, $\rho_{1}$, is varied for fixed resonance position, $\rho_{res}$ for off-axis low-field side $^{3}$He minority ICRH for different minority concentrations. Here the simulations with \textsc{Hagis} use the fast ion distribution from \textsc{Selfo}.}
\label{fig:ICRH_stability}
\end{figure}

Figure \ref{fig:ICRH_stability} shows the change in the potential energy of the mode arising due to the ICRH energetic ions as a function of the difference between the resonance radial location and the radius of the $q=1$ surface.
There is a clear narrow well in the potential energy when the RF resonance is just inside the rational surface, that is to say when the gradient of the distribution of energetic passing ions is strong and positive.
This narrow region ($\sim 2$cm) in which the sawteeth will be sensitive to the destabilising influence of the ICRH energetic ions implies that real-time control will be required in order that the resonance location be held in the right location with respect to the $q=1$ surface, though this is expected to be available between 40-55MHz in ITER with requisite latency \cite{Lamalle}.
Despite the fact that the power absorbed by the minority species increases with the concentration \cite{Chapman2011}, the strongest effect on mode stability is for a $^{3}$He concentration of only 1\%.
When there is too much $^{3}$He, the energy of the particles in the tail of the distribution becomes too low to have a strong effect on the kink mode whereas too little $^{3}$He means that the absorbed power is low and the broader distribution function leads to increased fast ion losses.
This strong sensitivity to minority concentration introduces a risk in using ICRH as a sawtooth control tool since accurate control of the minority concentration and radial profile is difficult to achieve.

\begin{figure}
\begin{center}
\includegraphics[width=0.7\textwidth,viewport = 0 0 600 350,clip]{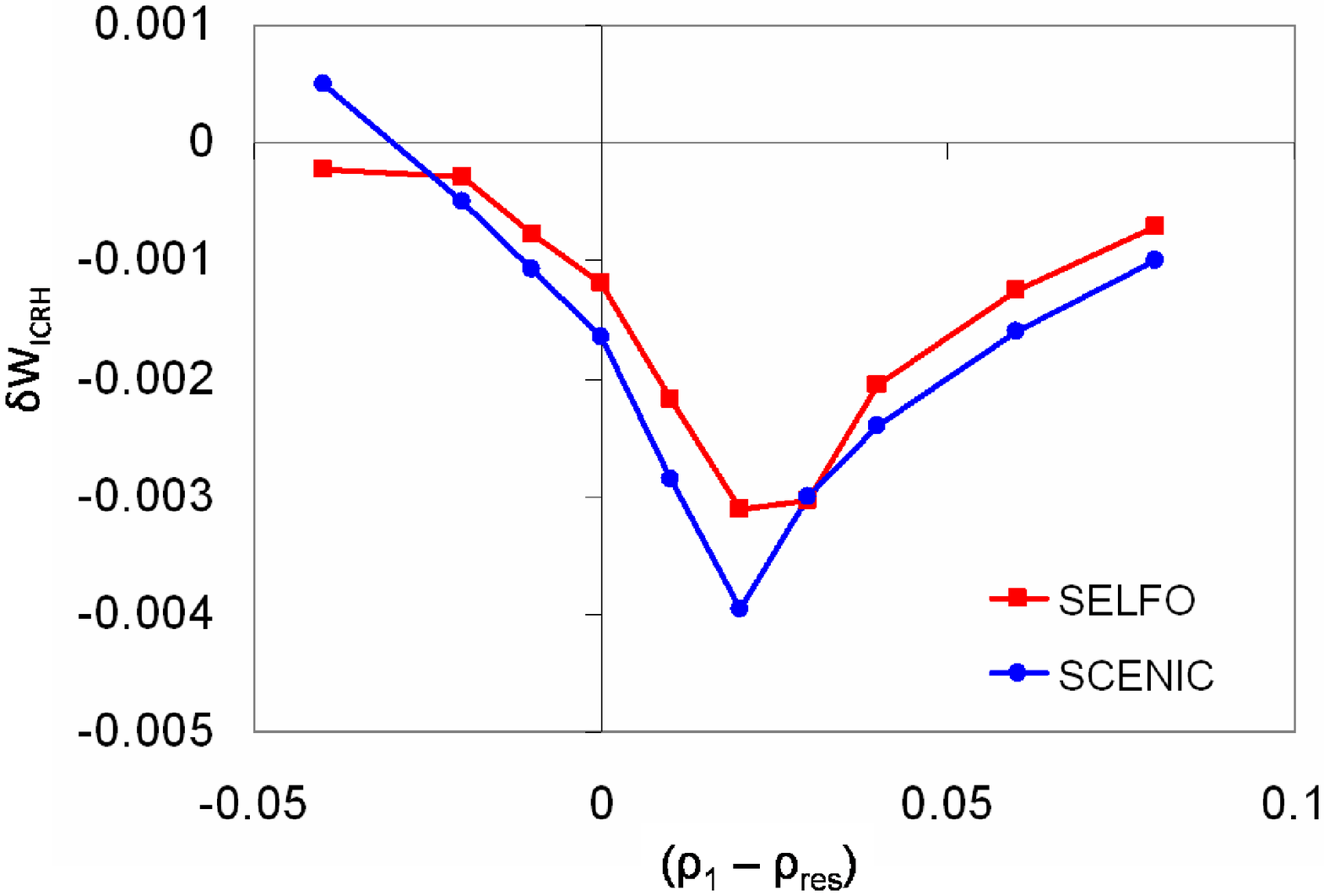}
\end{center}
\caption{A comparison of $\delta W_{ICRH}$ predicted by \textsc{Hagis} as $\rho_{1}$ is varied for fixed ICRH resonance position using distributions of markers from both \textsc{Selfo} and \textsc{Scenic} for a case with 0.5\% $^{3}$He, -90$^{\circ}$ antenna phasing and 20MW injected power with the resonance at $R=7.284$m.}
\label{fig:selfoscenic}
\end{figure} 

\begin{figure}
\hspace{-0.5cm}
\begin{minipage}{0.48\textwidth} 
\begin{center}
\includegraphics[width=\textwidth,viewport = 0 5 580 390,clip]{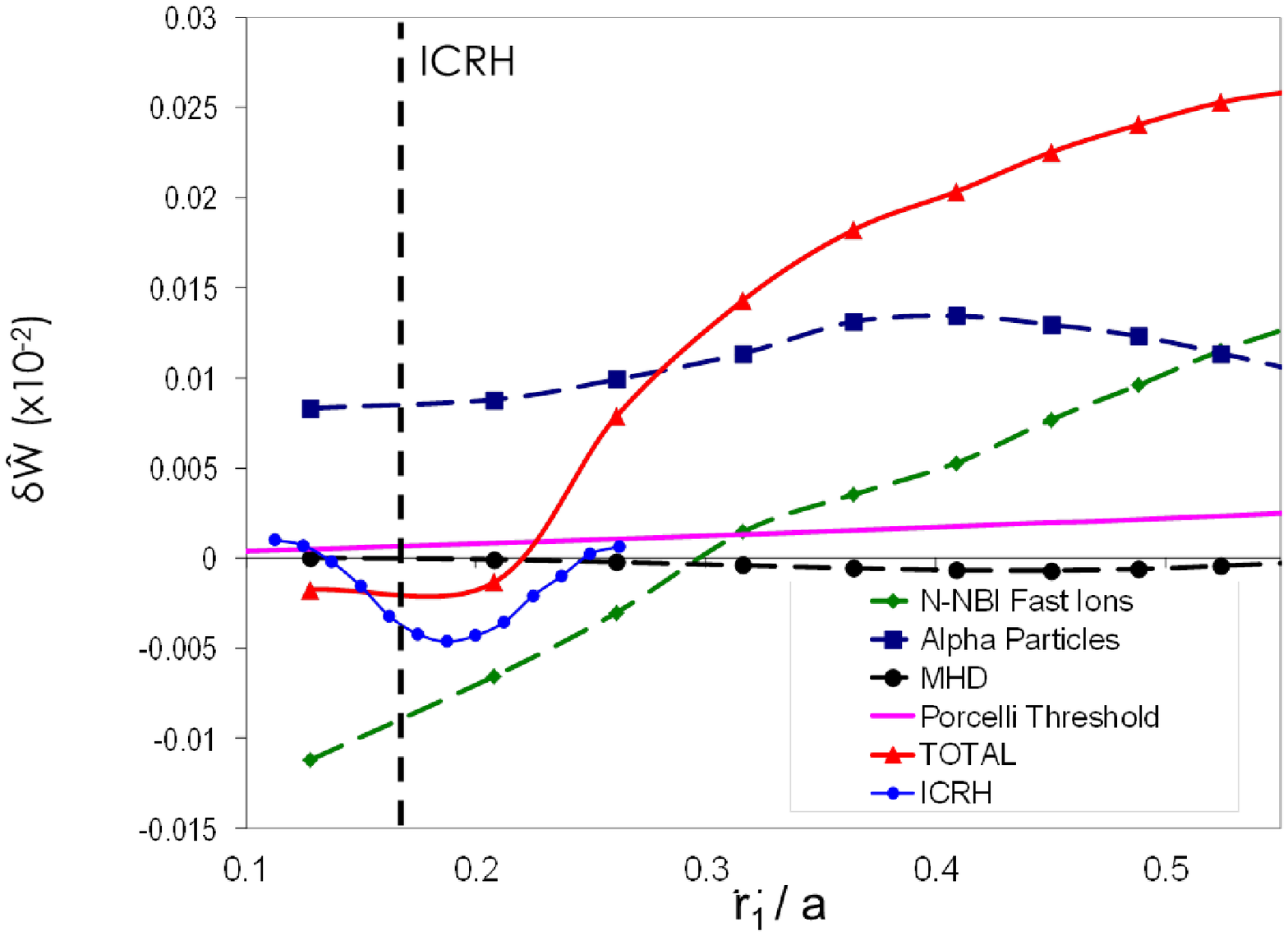}
\end{center}
\end{minipage}
\hspace{0.5cm} 
\begin{minipage}{0.48\textwidth}
\begin{center}
\includegraphics[width=\textwidth,viewport = 0 5 450 260,clip]{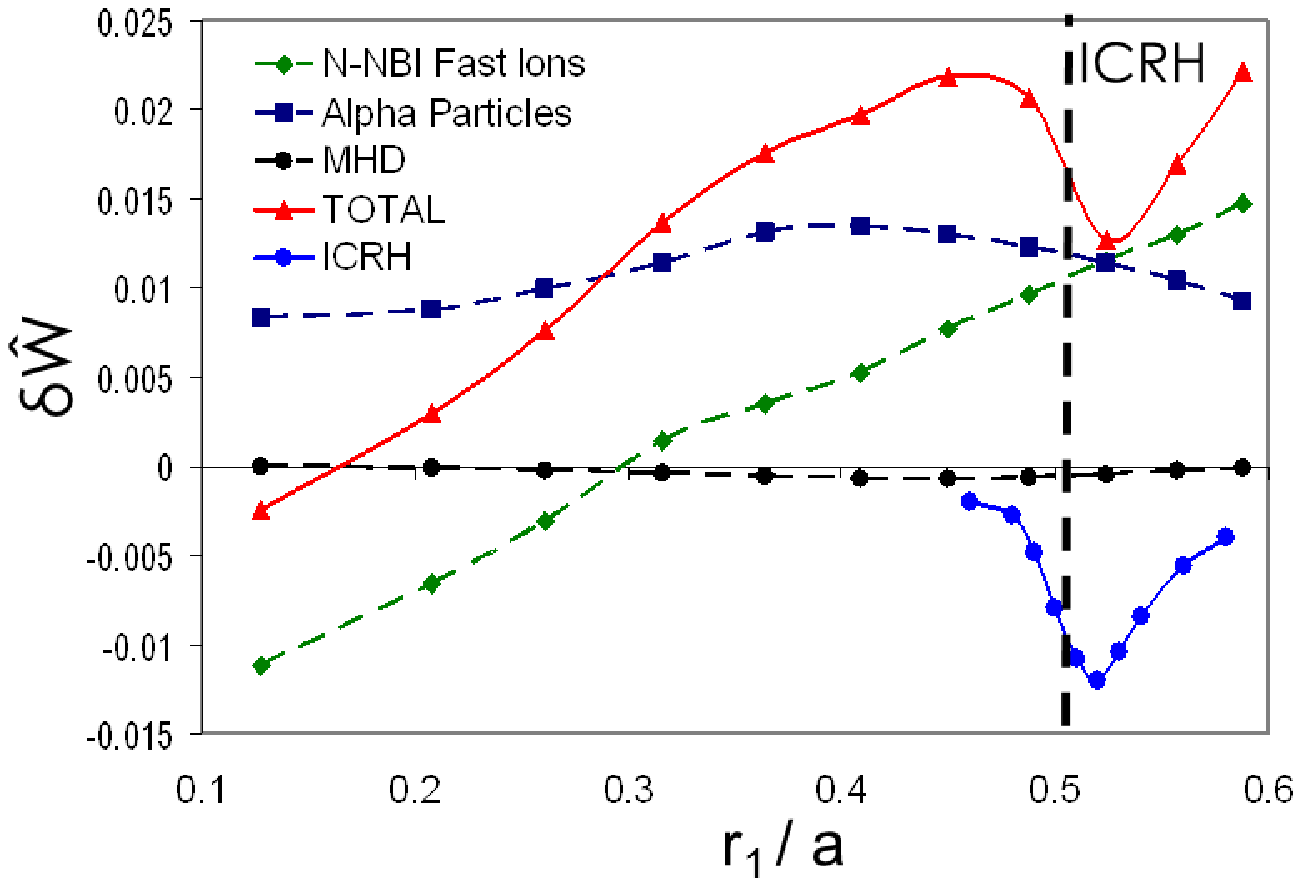}
\end{center}
\end{minipage}
\caption{The potential energy of the mode in the presence of all the competing fast ion distributions when (left) the ICRH is at $r=0.32$m, $f_{ICRH}=48.58$MHz and (right) the ICRH is at mid-radius, $f_{ICRH}=50$MHz. It is seen that ICRH significantly destabilises the mode, despite the strongly stabilising contribution from the $\alpha$'s and the NNBI ions. Here, the 20MW of ICRH is simulated by \textsc{Selfo}, the 33MW of off-axis NNBI with \textsc{Ascot} and the alphas with \textsc{Ascot}. The ICRH deposition is held fixed, as marked by the dashed vertical line, whilst the position of $q=1$ is varied.}
\label{fig:fastion_stability}
\end{figure}

A comparison of $\delta W_{ICRH}$ when the distribution of markers is taken from both \textsc{Selfo} or \textsc{Scenic} is shown in figure \ref{fig:selfoscenic} for the case with 0.5\% $^{3}$He, -90$^{\circ}$ antenna phasing and 20MW injected power with the resonance at $r=0.32$m (f=48.58MHz in \textsc{Selfo} and f=48.9MHz in \textsc{Scenic}).
It is clear that slightly stronger destabilisation is observed using the \textsc{Scenic} distribution, though in general the agreement is good.
This could be due to the inclusion of the shaping effects.
The main purpose of this comparison, though, is to mitigate the risk in the modelling uncertainty by taking the predictions from the least favourable result, in this case, the \textsc{Selfo} distribution.
The influence of the ICRH fast ions compared to the stabilising effect of the alpha particles and NNBI distributions is shown in figure \ref{fig:fastion_stability} for the case when the ICRH resonance is at $r=0.32$m ($f_{ICRH}=48.9\textrm{MHz}$) and when it is at $r=0.43m$ ($f_{ICRH}=50\textrm{MHz}$).
In these simulations the $q=1$ surface is moved by changing the equilibrium rather than re-simulating the fast ion distribution for different resonance locations.
It is evident that the mid-radius ICRH fast ions, despite the poor power absorption and low energy tails, retain a strongly destabilising influence, comparable to the magnitude of stabilisation afforded by the alphas or the NBI heating.
Whilst the power absorption is better when the resonance layer is nearer to the axis, resulting in improved core heating, the passing fast ions are only destabilising when the radial location of the $q=1$ surface is inside $\rho=0.2$.
These simulations are for 1\% $^{3}$He concentration and +90$^{\circ}$ phasing of the antenna, though the $-90^{\circ}$ phasing gives similar results, with a slightly diminished destabilisation.
The fact that the ICRH is able to completely negate the stabilising term from the presence of the $\alpha$ population is significant and important, and makes the ICRH an essential part of the portfolio of control tools in ITER.

Having reduced the risk in uncertainty of the ICRH fast ion distribution by utilising independent RF wave field codes, the largest residual uncertainty in this modelling is the location of the $q=1$ surface.
The ITER baseline scenario designed using \textsc{Astra} transport simulations suggests that the $q=1$ surface will approach mid-radius.
However, the $q$-profile has a wide region of very low shear in the core, meaning that a small change in $q_{0}$ can significantly affect the radial location of the rational surface.
Figure \ref{fig:fastion_stability} also shows that if the $q=1$ surface could be maintained closer to the magnetic axis, sawtooth control would be significantly easier to achieve, since the alphas would be less stabilising, the NNBI would be less stabilising, and could even be used as a destabilising control tool in the most off-axis orientation, and the control and flexibility afforded by the ICRH would be increased. 
Furthermore, the ECCD used to control the sawteeth would be closer to the plasma core, and so have the dual benefit of heating the plasma, hence affording a potential reduction in other auxiliary heating power and subsequent increase in $Q$.
This may be possible with early heating to delay the current penetration into the core, as regularly employed on JET, and then deliberate sawtooth destabilisation to mediate the $q$-profile once the $q=1$ surface enters.

\begin{figure}
\begin{center}
\includegraphics[width=0.7\textwidth,viewport = 0 5 580 390,clip]{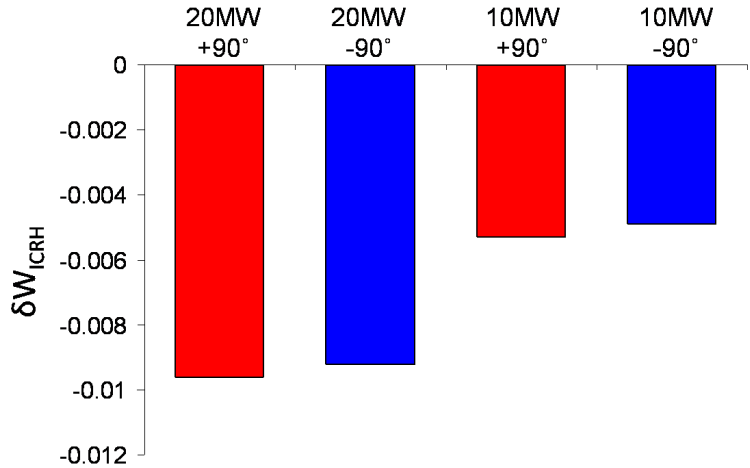}
\end{center}
\caption{The potential energy of the mode in the presence of 10 or 20MW of off-axis ICRH with $\pm 90^{\circ}$ antenna phasings and $\rho_{1}-\rho_{res}=0.02$ as predicted by \textsc{Hagis} using distributions from \textsc{Selfo}.}
\label{fig:powerscan}
\end{figure} 

\begin{figure}
\begin{center}
\includegraphics[width=0.7\textwidth,viewport = 0 5 400 300,clip]{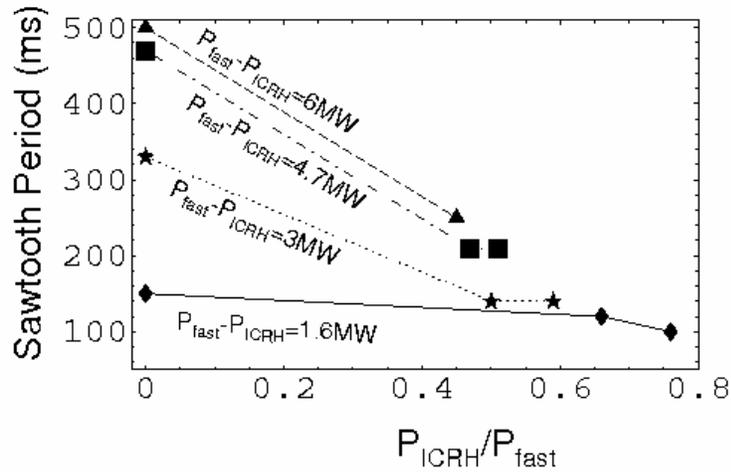}
\end{center}
\caption{The sawtooth period with respect to the injected $^{3}$He minority ICRH power for different total auxiliary input power in JET.}
\label{fig:icrh_power_jet}
\end{figure}

Since 20MW of ICRH power is unlikely to be dedicated for sawtooth control, lower ICRH powers have also been considered.
\textsc{Selfo} simulations of the ICRH fast ion population from 10MW of ICRH with 1\% $^{3}$He minority concentration, $\pm$90$^{\circ}$ antenna phasing and $f_{ICRH}=48.6$MHz have been performed and used as input for \textsc{Hagis}.
The change in the potential energy of the mode at the optimal relative position of $q=1$ with respect to the ICRH resonance location scales more favourably than linearly at reduced power.
The $\delta W_{ICRH}$ when $\rho_{1}-\rho_{res}=0.02$ (the optimal resonance position as seen in figure \ref{fig:ICRH_stability}) is shown in figure \ref{fig:powerscan} for both $\pm 90^{\circ}$ antenna phasings for 20MW and 10MW injected power.
Whilst it is not possible to infer the sawtooth period resultant from the ICRH application from this linear modelling, it is clear that 10MW of RF heating does significantly destabilise the kink mode, meaning that it is likely to be useful as an ancillary control actuator, even with half the available power.
This is supported by empirical evidence from recent JET experiments demonstrating sawtooth control with $^{3}$He minority schemes \cite{Graves2010}.
Figure \ref{fig:icrh_power_jet} shows the sawtooth period with respect to the ICRH power normalised to the total auxiliary power (NBI+RF) for the series of pulses described in reference \cite{Graves2010}. 
Each curve is for different injected NBI power.  
This can be considered congruent to including alphas where the fixed quantity is $P_{NBI}+P_{\alpha}=P_{fast}-P_{ICRH}$.
Whilst there is initially a strong monotonic reduction of the sawtooth period as the ICRH power is increased for all different background fast ion stabilisation levels, as the ICRH power continues to increase, the destabilisation does not increase.
This happens as the saturated sawtooth period is close to that of Ohmic sawteeth, or at least Ohmic sawteeth with modified energy and resistive diffusion time due to the heating effect of the ICRH.
Whatever level of ICRH power available in ITER for sawtooth control, it is likely to have a significant beneficial contribution to make in reducing the sawtooth period.

\section{Sawtooth Stabilisation} \label{sec:stabilisation}

If it proves that a combination of $>10$MW of ICRH power on top of the primary actuator of 13.3MW of ECCD from the equatorial launcher is insufficient for successful sawtooth control, then it is important that an alternative solution is provided within the design of the heating and current drive facilities available on ITER.
It is worth noting that in D-T plasmas in JET the best performance was achieved in the transient phases during which sawteeth were avoided \cite{Nave2002}.
Consequently, a sawtooth stabilisation scenario has been envisioned, whereby the natural sawtooth period is deliberately lengthened, and the (very probable) NTM that ensues at the crash is pre-emptively stabilised before it reaches its saturated width.
This was considered the most desirable route to sawtooth amelioration in the original ITER Physics Basis \cite{IPB2}, and was only superseded by destabilising control tools as anxiety grew about the ramifications of triggering performance-degrading NTMs and due to the need for frequent expulsion of the on-axis accumulation of higher-$Z$ impurities that would otherwise cause degradation of energy confinement due to impurity radiation. 

Long sawtooth periods are naturally achieved by applying early heating during the current ramp-up phase to increase the conductivity and so slow down the current penetration.
Combining this with achieving early ignition will further stabilise the sawteeth due to the $\alpha$ particle stabilisation.
ICRH could then be used as an ancillary control tool, with core heating providing a further population of strongly stabilising fast ions.
Furthermore, in order to meet the $Q=10$ goal of ITER baseline scenario, it is desirable to turn off the ECRH power whenever it is not being actively used for mode control, or to use it for core heating and reduce the NBI/ICRH power, and so maximise $Q$.
Thus, rather than being constantly required to modify the shear at $q=1$, an alternative can be foreseen whereby fast ions are used to deliberately stabilise the sawteeth, and before each crash the ECCD is pre-emptively applied near the $q=2$ surface to stabilise the ensuing NTM \cite{Sauter2010,LaHaye2009}.
Provided the seed island is sufficiently small ($\sim 7$cm), all models for ECCD island suppression indicate that 13.3MW from the upper launchers can fully suppress the mode within 6s \cite{Sauter2010}.
It is important to apply the ECCD pre-emptively in order to tackle the NTM when the island width is relatively small, thus requiring much less time of applied ECCD in order to fully suppress the mode, and therefore having a less deleterious effect on fusion gain.
However, it is worth noting that the seed island for NTM growth can be created at large island size at the sawtooth crash as sometimes observed in JET \cite{Sauter,Graves2010}, meaning that the pre-emptive ECCD is only useful if it modifies the dynamics of the island seeding during the sawtooth crash.

It is anticipated that by using either counter-CD in the core, off-axis co-ECCD or on-axis ICRH to maximise the sawtooth period, transient (though this could be $>100$s) periods of very high $Q$ could be attained.
It could be that if the $\alpha$ stabilisation proves to be stronger than anticipated, and the natural sawtooth period is $>100$s, then control by deliberate stabilisation, followed by provocation of a crash through dropping auxilliary heating coupled with simultaneous pre-emptive application of ECCD near the $q=2$ surface to suppress the subsequent NTM growth, could be the best route to long-periods of excellent performance.
This has been demonstrated in TCV where sawtooth pacing using modulated core ECCD coupled with pre-emptive NTM avoidance by ECCD at $q=3/2$ has avoided the triggering of NTMs \cite{Goodman,Felici}.
It should be noted that deliberate stabilisation of the sawteeth using either ICRH or core counter-ECCD/off-axis co-ECCD will require real-time control, as discussed in section \ref{sec:ECCD}.
Whilst sawtooth stabilisation may even seem advantageous for maximising $Q$ due to the reduced application time of the control actuators, the removal of frequent sawteeth from the baseline scenario means an alternative strategy is required to reverse the on-axis accumulation of higher-$Z$ impurities that would otherwise cause degradation of energy confinement due to impurity radiation.
Although this has been demonstrated using core ECH in ASDEX Upgrade \cite{Dux,Neu}, its application for $He$-ash removal is untested, and the avoidance of potentially disruptive NTMs (even with pre-emptive ECCD) is preferable, so sawtooth destabilisation remains the optimal solution for ITER.

\section{Conclusions} \label{sec:conclusions}

An empirical scaling of the sawtooth period that will trigger an NTM in ITER suggests that the ``natural'' sawtooth period predicted by transport modelling is approximately at the threshold for NTM seeding.
Whilst this means that active sawtooth control is essential, it suggests that sufficient control can be achieved through a relatively small reduction in the sawtooth period.
Transport modelling coupled to ray-tracing predictions and using the linear stability thresholds for sawtooth onset suggests that 13MW of ECCD from the equatorial launcher could be sufficient to reduce the sawtooth period by $\sim$30\%, and this being the case, dropping it below the NTM triggering threshold.
This modelling is predicated upon choosing a natural sawtooth period of 40s; should the stabilising contribution from the fusion-born alpha particles and on-axis NBI injection prove to give rise to a significantly longer natural sawtooth period, the ability of the ECCD to control sawteeth will be diminished.
There are naturally large uncertainties associated with this modelling, and it is prudent to plan to use more than one control actuator in order to reduce this risk.
Consequently, it is recommended that $>10$MW of ICRH at $\sim 47$MHz (with real-time feedback) just inside $q=1$ is also reserved for sawtooth control.
The largest uncertainty in the modelling of the effect of the fast ions is the position of the $q=1$ surface.
If the $q=1$ surface could be maintained closer to the magnetic axis, sawtooth control would be significantly easier to achieve, since both the alphas and the beam-induced fast ions would be less stabilising.
Furthermore, the ICRH and ECCD used to control the sawteeth would be further towards the plasma core, thus heating in the good confinement region and so affording a potential reduction in other auxiliary heating power and subsequent increase in $Q$.
Finally, should active sawtooth destabilisation prove to be unattainable due to unexpectedly large stabilising contribution from the $\alpha$ particles, plant availability or inefficiency in power absorption or current drive, then there is a viable alternative strategy relying upon sawtooth stabilisation coupled with pre-emptive NTM suppression, which would provide long periods of good performance.
The power requirements for the necessary degree of sawtooth control using either destabilisation and stabilisation schemes are expected to be within the specification of anticipated ICRH and ECRH heating in ITER, provided the requisite power can be dedicated to sawtooth control.
It is worth reiterating that the results presented in sections \ref{sec:empirical} and \ref{sec:exp} are derived from present day experiments in the absence of alpha particles. The alpha particles are found to significantly stabilise the internal kink mode, as shown in section \ref{sec:ICRH}, which means there is a relatively high uncertainty in the modelling predictions at high alpha fraction, that is to say at high $Q$.  

\vspace{1cm}
\noindent \textbf{Acknowledgements}

\noindent This work was conducted under the auspices of the ITPA MHD Stability Topical Group.
This work was partly funded by the United Kingdom Engineering and Physical Sciences Research Council under grant EP/I501045, the European Communities under the contract of Association between EURATOM and CCFE and supported in part by the Swiss National Science Foundation and US Department of Energy under DE-FC02-04ER54698 and DE-AC52-07NA27344. The views and opinions expressed herein do not necessarily reflect those of the European Commission.




\vspace{1cm}
\begin{thebibliography}{99} 
\bibitem{Casper}
T Casper \emph{et al} {\it 23rd IAEA Fusion Energy Conference, Daejon} ITR/P1-19 (2010)
\bibitem{Porcelli1996}
F Porcelli, D Boucher and M Rosenbluth, {\it Plasma Phys. Control. Fusion} \textbf{38} 2163 (1996)
\bibitem{ChapmanEPS}
IT Chapman et al, {\it Plasma Phys. Control. Fusion} \textbf{49} B385 (2007)
\bibitem{Sauter}
O Sauter et al, {\it Phys. Rev. Lett.} \textbf{88} 105001 (2002)
\bibitem{Buttery}
RJ Buttery et al. in Fusion Energy 2004  (Proc. 20th Int. Conf. Vilamoura, 2004) (Vienna: IAEA) CD-ROM file EX/7-1 and http://www-naweb.iaea.org/napc/physics/fec/fec2004/datasets/index.html
\bibitem{Chapman2010}
IT Chapman \emph{et al} 2010 {\it Nucl Fusion} \textbf{50} 102001
\bibitem{ChapmanRev}
IT Chapman 2011 {\it Plasma Phys. Control. Fusion} \textbf{53} 003001
\bibitem{IPB2}
T C Hender et al. Progress in the ITER Physics Basis Chapter 3: MHD stability, operational limits and disruptions 2007 Nucl. Fusion \textbf{47} S128-S202
\bibitem{Graves2004}
JP Graves, {\it Phys. Rev. Lett.} \textbf{92} 185003 (2004)
\bibitem{Graves2009}
JP Graves et al, {\it Phys. Rev. Lett.} \textbf{102} 065005 (2009)
\bibitem{Chapman2007}
IT Chapman et al, {\it Phys. Plasmas} \textbf{14} 070703 (2007)
\bibitem{Chapman2008}
IT Chapman et al, {\it Plasma Phys. Control. Fusion} \textbf{50} 045006 (2008)
\bibitem{Chapman2009}
Chapman IT \emph{et al} 2009 {\it Nucl. Fusion} \textbf{49} 035006
\bibitem{GravesAPS}
JP Graves et al, 2009 {\it Phys. Plasmas} \textbf{17} 056118
\bibitem{Laxaback}
M Laxaback and T Hellsten, {\it Nucl. Fusion} \textbf{45} 1510 (2005)
\bibitem{Graves2012}
JP Graves \emph{et al} 2012 {\it Nat. Comms.} \textbf{3} 624
\bibitem{Graves2010}
JP Graves \emph{et al} 2010 {\it Nucl Fusion} \textbf{50} 052002
\bibitem{Carrera}
R Carrera, RD Hazeltine and M Kotschenreuther 1986 {\it Phys Fluids} \textbf{29} 899
\bibitem{Brennan}
DP Brennan et al 2003 {\it Phys. Plasmas} \textbf{10} 1643
\bibitem{Fitzpatrick}
R Fitzpatrick et al, 1995 {\it Phys. Plasmas} \textbf{2} 825
\bibitem{Wilson}
HR Wilson et al, 1996 {\it Phys. Plasmas} \textbf{3} 248
\bibitem{Smolyakov}
AI Smolyakov, A Hirose, E Lazzaro, GB Re and JD Callen 1995 {\it Phys Plasmas} \textbf{2} 1581
\bibitem{Kotschenreuther}
M Kotschenreuther, RD Hazeltine and PJ Morrison 1985 {\it Phys Fluids} \textbf{28} 294
\bibitem{Lutjens}
H L\"{u}tjens, JF Luciani and X Garbet 2001 {\it Phys Plasmas} \textbf{8} 4267
\bibitem{LaHaye2000}
R.J. La Haye et al., {\it Nucl Fusion} \textbf{40}, 53 (2000)
\bibitem{Buttery2008}
RJ Buttery et al 2008 {\it Phys. Plasmas} \textbf{15} 056115
\bibitem{Gerhardt}
S Gerhardt et al 2009 {\it Nucl. Fusion} \textbf{49} 032003
\bibitem{Gude}
A Gude, S Guenter and S Sesnic 1999 {\it Nucl. Fusion} \textbf{39} 127
\bibitem{Hegna}
CC Hegna, JD Callen and RJ La Haye 1999 {\it Phys. Plasmas} \textbf{6} 130
\bibitem{Nave}
MFF Nave et al 2003 {\it Nucl. Fusion} \textbf{43} 179
\bibitem{Reimerdes}
H Reimerdes et al 2002 {\it Phys. Rev. Lett.} \textbf{88} 105005
\bibitem{Maget}
P Maget \emph{et al} 2005 {\it Plasma Phys Control Fusion} \textbf{47} 357
\bibitem{Koslowski}
HR Koslowski \emph{et al} 2000 {\it Nucl. Fusion} \textbf{40} 821
\bibitem{Buttery2003}
RJ Buttery et al 2003 {\it Nucl. Fusion} \textbf{43} 69
\bibitem{Sauter1997}
O Sauter et al 1997 {\it Phys. Plasmas} \textbf{4} 1654
\bibitem{Belo}
PA Belo et al 2001 {\it in Proceedings of the 28th EPS Conference on Controlled Fusion and Plasma Physics, Portugal} P5.004
\bibitem{Coda}
S Coda et al 2007 {\it in Proceedings of the 34th EPS Conference on Controlled Fusion and Plasma Physics, Warsaw} P5.130
\bibitem{Buttery2004}
RJ Buttery et al, {\it Nucl. Fusion} \textbf{44} 678 (2004)
\bibitem{Jardin}
SC Jardin, MG Bell and N Pomphrey 1993 {\it Nucl. Fusion} \textbf{33} 371
\bibitem{Bateman1998}
G Bateman \emph{et al} 1998 {\it Phys. Plasmas} \textbf{5} 2355
\bibitem{Bateman} 
Bateman G, Nguyen CN, Kritz AH and Porcelli F 2006 {\it Phys. Plasmas} \textbf{13} 072505
\bibitem{Waltz}
R Waltz \emph{et al} 1997 {\it Phys Plasmas} \textbf{4} 2482
\bibitem{Onjun}
T Onjun and Y Pianroj 2009 {\it Nucl. Fusion} \textbf{49} 075003
\bibitem{Budny2008}
RV Budny \emph{et al} 2008 {\it Nucl. Fusion} \textbf{48} 075005
\bibitem{Martin}
YR Martin \emph{et al} 2008 {\it Journal of Physics: Conference Series} \textbf{123} 012033
\bibitem{Doyle}
EJ Doyle \emph{et al} Progress in the ITER Physics Basis Chapter 2: Plasma
confinement and transport 2007 {\it Nucl. Fusion} \textbf{47} S18
\bibitem{Lennholm}
M Lennholm et al, {\it Phys. Rev. Lett.} \textbf{102} 115004 (2009)
\bibitem{Lennholm2009}
Lennholm M \emph{et al} 2009 {\it Fus. Sci. Tech.} \textbf{55} 45 
\bibitem{Igochine}
Igochine V \emph{et al} 2011 {\it Plasma Phys. Control Fusion} \textbf{53} 022002
\bibitem{Muck}
M\"{u}ck A, Goodman TP, Maraschek M, Pereverez G, Ryter F and Zohm H 2005 {\it Plasma Phys. Control Fusion} \textbf{47} 1633
\bibitem{Isayama}
Isayama A \emph{et al} 2002 {\it J. Plasma Fus. Res. Series} \textbf{5} 324
\bibitem{ChapmanDIIID}
IT Chapman \emph{et al} 2012 {\it Nucl Fusion} \textbf{52} 063006
\bibitem{ChapmanAUG}
IT Chapman \emph{et al} 2012 {\it in prep Plasma Phys. Control. Fusion} ``Destabilisation of sawteeth in the presence of core energetic particles in ASDEX Upgrade''
\bibitem{Paley}
Paley JI \emph{et al} 2009 {\it Plasma Phys. Control. Fusion} \textbf{51} 055010
\bibitem{GoodmanPRL}
TP Goodman \emph{et al} 2011 {\it Phys Rev Lett} \textbf{106} 245002
\bibitem{Lauret}
M Lauret \emph{et al} 2012 {\it Nucl Fusion} \textbf{52} 062002
\bibitem{Witvoet}
G Witvoet \emph{et al} 2011 {\it Nucl Fusion} \textbf{51} 103043
\bibitem{Westerhof2002}
Westerhof E \emph{et al} 2002 {\it Nucl. Fusion} \textbf{42} 1324
\bibitem{Mayoral}
Mayoral ML \emph{et al} 2004 {\it Phys. Plasmas} \textbf{11} 2607
\bibitem{Eriksson2004}
Eriksson L-G \emph{et al} 2004 {\it Phys. Rev. Letters} \textbf{92} 235004
\bibitem{Eriksson2006}
Eriksson L-G \emph{et al} 2006 {\it Nucl. Fusion} \textbf{46} S951
\bibitem{Lennholm2011}
M Lennholm et al, 2011 {\it Nucl Fusion} \textbf{51} 073032
\bibitem{Budny1992}
RV Budny {\it Nucl. Fusion}  1992 {\it Nucl. Fusion} \textbf{32} 429
\bibitem{Heikennen}
J. A. Heikkinen and S. K. Sipila, {\it Phys Plasmas}, \textbf{2} (1995) 3724
\bibitem{Kurki}
T. Kurki-Suonio et al., {\it Nucl Fusion} \textbf{49} (2009) 095001
\bibitem{Suzuki}
S. Suzuki et al., {\it Plasma Physics and Controlled Fusion}, \textbf{40} (1998) 2097
\bibitem{Hedin}
J Hedin et al, {\it Nucl. Fusion} \textbf{42} 527 (2002)
\bibitem{Jucker}
M Jucker \emph{et al} 2011 {\it Comp. Phys. Comm.} \textbf{182}, 912
\bibitem{Jucker2}
M Jucker \emph{et al} 2011 {\it Plasma Phys. Control Fusion} \textbf{53} 054010
\bibitem{Pinches}
SD Pinches et al, {\it Comput. Phys. Commun.} \textbf{111} 133 (1998) 
\bibitem{Villard}
L. Villard, S. Brunner, and J. Vaclavik (1995) Nuclear Fusion 35 1173
\bibitem{Carlsson}
J. Carlsson, L.-G. Eriksson, and T. Hellsten (1997) Nuclear Fusion 37(6) 719
\bibitem{Stix}
Stix, Waves in Plasmas, p258
\bibitem{BudnyIAEA}
RV Budny et al. in Fusion Energy 2010 (Proc. 23rd Int. Conf. Daejeon, 2010) (Vienna: IAEA) CD-ROM file ITR/P1-29 and http://www-naweb.iaea.org/napc/physics/FEC/FEC2010/html/index.htm
\bibitem{Cooper}
WA Cooper \emph{et al} 2006 {\it Nucl Fusion} \textbf{46} 683
\bibitem{Popovich}
P Popovich, WA Cooper and L Villard 2006 {\it Comp. Phys. Comm.} \textbf{175} 250
\bibitem{Fischer}
O Fischer, WA Cooper, MY Isaev and L Villard 2002 {\it Nucl. Fusion} \textbf{42} 817
\bibitem{Cooper2007}
G. A. Cooper et al, {\it Phys Plasmas} \textbf{14} 102506 (2007)
\bibitem{Pereverzev}
G. V. Pereverzev, Max Planck - IPP Report 5/42, (1991)
\bibitem{Porcelli1991}
Porcelli F 1991 {\it Plasma Phys. Control. Fusion} \textbf{33} 1601
\bibitem{Kruskal}
Kruskal M and Oberman C 1958 {\it Phys. Fluids} \textbf{1} 275
\bibitem{SauterVarenna}
Sauter O \emph{et al} 1998 {\it Theory of Fusion Plasmas, Proc Joint Varenna-Lausanne International Workshop, Varenna} (AIP) p403
\bibitem{Bussac}
Bussac MN \emph{et al} 1975 {\it Phys. Rev. Lett.} \textbf{35} 1638
\bibitem{Huysmans}
Huysmans GTA \emph{et al}, 1991 \emph{Proc CP90 Conf on Comp Phys} p.371
\bibitem{ChapmanMishka}
Chapman IT, Huysmans GTA, Mikhailovskii AB and Sharapov SE 2006 {\it Phys. Plasmas} \textbf{13} 062511
\bibitem{ChapmanPoP}
Chapman IT \emph{et al} 2009 {\it Phys. Plasmas} \textbf{16} 072506
\bibitem{Nave2002}
Nave MFF \emph{et al} 2002 {\it Nucl. Fusion} \textbf{42} 281
\bibitem{Chapman2011}
Chapman IT \emph{et al} 2011 {\it Plasma Phys. Control. Fusion} \textbf{53} 124003
\bibitem{ITER}
ITER Technical Basis for Final Design 2001 {\it ITER Documentation Series} \#24 (Vienna: IAEA) Chapter 2.5, Page 2
\bibitem{Budny2002}
Budny RV 2002 {\it Nucl. Fusion} \textbf{42} 1383
\bibitem{Jaeger}
EF Jaeger \emph{et al} 2006 {\it Phys Plasmas} \textbf{13} 056101
\bibitem{Harvey}
R Harvey and M McCoy 1992 {\it Proc IAEA Technical Committee Meeting on Numerical Modelling of Plasmas, Montreal, Canada} Vienna:IAEA
\bibitem{Omori}
T Omori \emph{et al} 2011 {\it Fus. Eng. Des.} \textbf{86} 951
\bibitem{Ramponi}
G Ramponi \emph{et al} 2008 {\it Nucl. Fusion} \textbf{48} 054012
\bibitem{Sauter1999}
O Sauter et al, Phys, Plasmas 6, 2834 (1999)
\bibitem{Sauter1999b}
O, Sauter et al, Phys, Plasmas 6, 5140 (2002)
\bibitem{Gray}
D Farina, Fusion Science and Technology 52, 154 (2007)
\bibitem{GravesFST}
JP Graves \emph{et al} 2011 {\it Fus. Sci. Tech.} \textbf{59} 539
\bibitem{Zucca}
C Zucca \emph{et al} 2008 {\it Theory of Fusion Plasmas, Joint Varenna-Lausanne Theory Conference} \textbf{1069} 361
\bibitem{Zucca2}
C Zucca 2009 {\it PhD Thesis No 4360, EPFL, Lausanne} ``Modeling and control of the current density profile in tokamaks and its relation to electron transport''
\bibitem{Kirneva}
NA Kirneva \emph{et al} 2012 {\it Plasma Phys Control Fusion} \textbf{54} 015011
\bibitem{Lamalle}
P Lamalle, ITER Organisation 2010 {\it Private Communication}
\bibitem{Sauter2010}
O Sauter \emph{et al} 2010 {\it Plasma Phys. Control. Fusion} \textbf{52} 025002\bibitem{LaHaye2009}
RJ La Haye et al 2009 {\it Nucl Fusion} \textbf{49} 045005
\bibitem{Goodman}
TP Goodman \emph{et al} 2012 {\it European Physical Society Conf on Plasma Phys, Stockholm, Sweden} P2.029 ocs.ciemat.es/epsicpp2012pap/pdf/P2.029.pdf 
\bibitem{Felici}
F Felici \emph{et al}, 2012 {\it Nucl Fusion} \textbf{52} 074001
\bibitem{Dux}
Dux R \emph{et al}, 2003 {\it Journal of Nuclear Materials} \textbf{313} 1150
\bibitem{Neu}
Neu R \emph{et al}, 2002 {\it Plasma Phys. Control. Fusion} \textbf{44}  811



\end{thebibliography}
\end{document}